\documentclass[notitlepage,onecolumn,amsmath,amssymb,aps]{revtex4-2}
\usepackage{graphicx}
\usepackage{bm}
\usepackage[caption=false]{subfig}
\usepackage{setspace}
\usepackage[colorinlistoftodos]{todonotes}
\usepackage{hyperref}
\usepackage[version=4]{mhchem}
\usepackage{xr}
\usepackage{lineno}
\usepackage{soul}

\makeatletter
\newcommand*{\addFileDependency}[1]{
  \typeout{(#1)}
  \@addtofilelist{#1}
  \IfFileExists{#1}{}{\typeout{No file #1.}}
}
\makeatother

\newcommand*{\myexternaldocument}[1]{%
    \externaldocument{#1}%
    \addFileDependency{#1.aux}%
}

\myexternaldocument{supplement-ref}

\begin{document}

\title{Communication network model of the immune system identifies the impact of interactions with SARS-CoV-2 proteins}

\author{Swarnavo Sarkar}
\email{swarnavo.sarkar@nist.gov}
\email{swarnavo.sarkars@gmail.com}
\affiliation{National Institute of Standards and Technology, Gaithersburg, MD 20899, USA}

\begin{abstract}  
Interactions between SARS-CoV-2 and human proteins (SARS-CoV-2 PPIs) cause information transfer through biochemical pathways that contribute to the immunopathology of COVID-19. Here, we present a communication network model of the immune system to compute the information transferred by the viral proteins using the available SARS-CoV-2 PPIs data. The amount of transferred information depends on the reference state of the immune system, or the state without SARS-CoV-2 PPIs, and can quantify how many variables of the immune system are controlled by the viral proteins. The information received by the immune system proteins from the viral proteins is useful to identify the biological processes (BPs) susceptible to dysregulation, and also to estimate the duration of viral PPIs necessary for the dysregulation to occur. We found that computing the drop in information from viral PPIs due to drugs provides a direct measure for the efficacy of therapies.
\end{abstract}

\maketitle

\begin{spacing}{2.0}

\section*{\label{se: introduction}Introduction}

COVID-19, the disease caused by severe acute respiratory syndrome coronavirus 2 (SARS-CoV-2), has multiple routes for harming human biology. One mechanism is through protein-protein interactions (PPIs) between SARS-CoV-2 and human proteins (SARS-CoV-2 PPIs) \cite{gordon2020sars}. The viral PPIs can alter biochemical functions of human proteins, the effect of which can propagate through the cascade of biochemical reactions informing the abundance of other proteins that do not directly interact with the viral proteins. The information transfer caused by SARS-CoV-2 PPIs may elucidate some of the more complex consequences of COVID-19, like long lasting immunological dysfunction \cite{phetsouphanh2022immunological,sudre2021attributes,nalbandian2021post}, dysregulation of myeloid response \cite{vabret2020immunology}, lymphopenia \cite{azkur2020immune,laing2020dynamic}, and more. For a comprehensive assessment of COVID-19, we need to quantify the information about SARS-CoV-2 PPIs that is communicated to the immune system proteins and the sensitivity of biological processes (BPs) to the amount of transferred information. The fundamental framework for communication is information theory \cite{cover1999elements}, which has been used to quantify the efficacy of biochemical signaling \cite{cheong2011information,bialek2005physical,suderman2017fundamental}. Information theory uses entropy to measure the dependence between stochastic variables, and all biochemical entities are stochastic at the cellular level. By bringing the fundamental information-theoretic model of biochemical signaling to the immune system, we can computationally determine the ramifications of COVID-19 and other infections. Computing the information transfer in the immune system due to viral PPIs also provides a framework for \textit{in silico} evaluation of therapies \cite{cava2020silico}.  

Here, we developed a communication network model of the immune system using pathways and interaction data to compute the information transferred by viral PPIs, and to subsequently identify potentially dysregulated BPs and efficacy of immunotherapies. We modeled biochemical reactions and human protein-protein interactions as channels transferring information \cite{cover1999elements,shannon1948mathematical} between the physical entities (proteins, complexes, metabolites, \textit{etc.}) that are part of the immune system. In this case study, the SARS-CoV-2 protein interaction map \cite{gordon2020sars} provides the sources of information in the immune system communication network. To determine the information transferred to the immune system proteins, we use an algorithm that is sensitive to the inter-connectivity of the immune system network \cite{sarkar2021information}, resulting in the interference of information reaching a physical entity from multiple pathways. We found that the information transfer from SARS-CoV-2 PPIs depends on the reference state of the immune network (\textit{i.e.}, the state in the absence of SARS-CoV-2 PPIs), which subsequently determines the BPs that can be dysregulated due to communication. Using the large deviations theory \cite{cover1999elements,varadhan1984large}, we estimate when a BP becomes significant for dysregulation due to communication from SARS-CoV-2 PPIs. Furthermore, we directly compute the reduction in information from viral PPIs due to candidate therapies, providing a new measure to rank the efficacy of drugs. 

\section*{\label{sec: network model}Immune system communication network}

To construct the communication network model of the immune system, we modeled the immune system as a graph $G=(V,E)$ made of nodes, $V$, and edges, $E$. We obtained the nodes and the edges from two databases: the Reactome database for the physical entities participating in the immune system and the reactions interconnecting them \cite{jassal2020reactome}, and the Human Reference Interactome database for the pairwise interactions between proteins \cite{luck2020reference}. The physical entities form the nodes of the network, whereas the reactions and the interactions contribute as edges (\ref{sec: network construction}). The immune system communication network consists of $|V|=3011$ nodes, with 1141 protein nodes (Fig. \ref{fig: immune network}, Table \ref{table: node composition}). We represented the state of each node as a Bernoulli random variable: if $v_i \in V$ is a network node, then the probabilistic state of the node is $\{P(v_i=1),P(v_i=0)\}$, where the abundance (no. of molecules) of $v_i$ is coarse-grained into a binary variable, 1 (abundant) and 0 (deficient). The state of the entire immune system network is a $2|V|$-dimensional vector $X$, where the $i^{th}$ 2-dimensional block of $X$ contains the state of $v_i$. The reactions and interactions connect pairs of nodes with input-output relationships (\ref{sec: network construction}). 

To simulate the flow of information in the immune system, we constructed a global transition matrix to iteratively update the network state due to communication. We modeled each edge as a binary information channel, with a 2-by-2 transition matrix $P(output|input)$, which communicates the state of the input to the output (\ref{sec: channel model}). Then we used the connectivity of the network $G$, and unbiased interference of information, to assemble the transition matrix of individual edges into the transition matrix for the entire network $C_G$ (\ref{sec: channel model}) \cite{sarkar2021information}. Information transfer through the network updates the state of the network as
\begin{equation}
\label{eq: iterative info flow}
X^{(k+1)} = C_G X^{(k)}
\end{equation}
where $X^{(k)}$ is the network state after $k$ steps of information propagation. If the initial state is $X^{(0)}$, then after $k$ steps of information transfer the state is $X^{(k)}=C_G^k X^{(0)}$. The stationary state of the network satisfies $X=C_G X$. There are 89 source nodes in the network (\textit{i.e.}, nodes that only send but do not receive information), the state of these nodes determine the reference state of the immune system communication network in the absence of other information sources, like SARS-CoV-2 PPIs. Current limitations of the communication network model include, incompleteness and errors in the pathways and interactions data, a binary channel to model the reactions and interactions, and unbiased interference of information. Proteomic, metabolomic, and rate constants data for human cells can provide data-driven parameters for channel bit size and interference bias. We expect to progressively integrate such data in the future versions of the model.

\section*{\label{sec: relH} Information transferred by SARS-CoV-2 PPIs}

To evaluate the information transferred by SARS-CoV-2 PPIs, we assumed that interactions with SARS-CoV-2 proteins diminish the function of human proteins. So, the impact of SARS-CoV-2 PPIs is a reduction in the \textit{functional} or the \textit{effective} abundance of the associated human proteins. For each of the proteins that directly interact with SARS-CoV-2 proteins (SARS-CoV-2 PPI node), we constrained the state of that protein node to $\{0,1\}$, or effectively to low abundance. We need experimental studies to assess the exact decline, if at all, in a protein's function due to SARS-CoV-2 PPIs (as available for interactions with the ORF9b protein \cite{thorne2021evolution}). However, we can computationally test all possible impact on the protein function, \textit{i.e.}, maximum decline in function (constraint $\{0,1\}$), no effect on the function (no constraint), or maximum increase in function (constraint $\{1,0\}$), or any intermediate constraint between $\{1,0\}$ and $\{0,1\}$. We used the constraint $\{0,1\}$ to compute the consequence of maximum decline in a protein's function due to SARS-CoV-2 PPIs.

\begin{figure*}[!ht]
    \begin{minipage}{0.49\textwidth}
    \subfloat[\label{fig: immune network}]{%
    \includegraphics[width=0.99\textwidth]{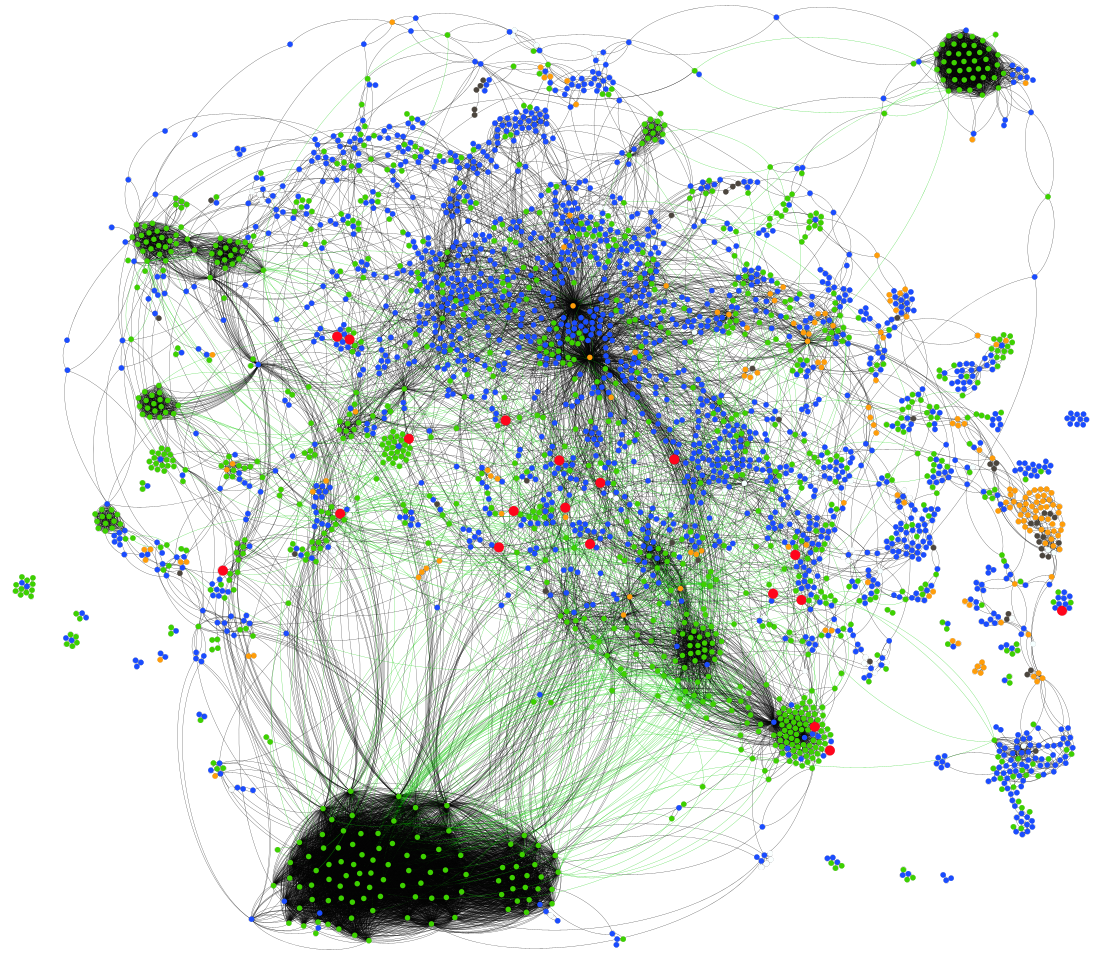}
    }
    \end{minipage}
    \begin{minipage}{0.49\textwidth}
    \subfloat[\label{fig: all protocols}]{%
    \includegraphics[width=0.99\textwidth]{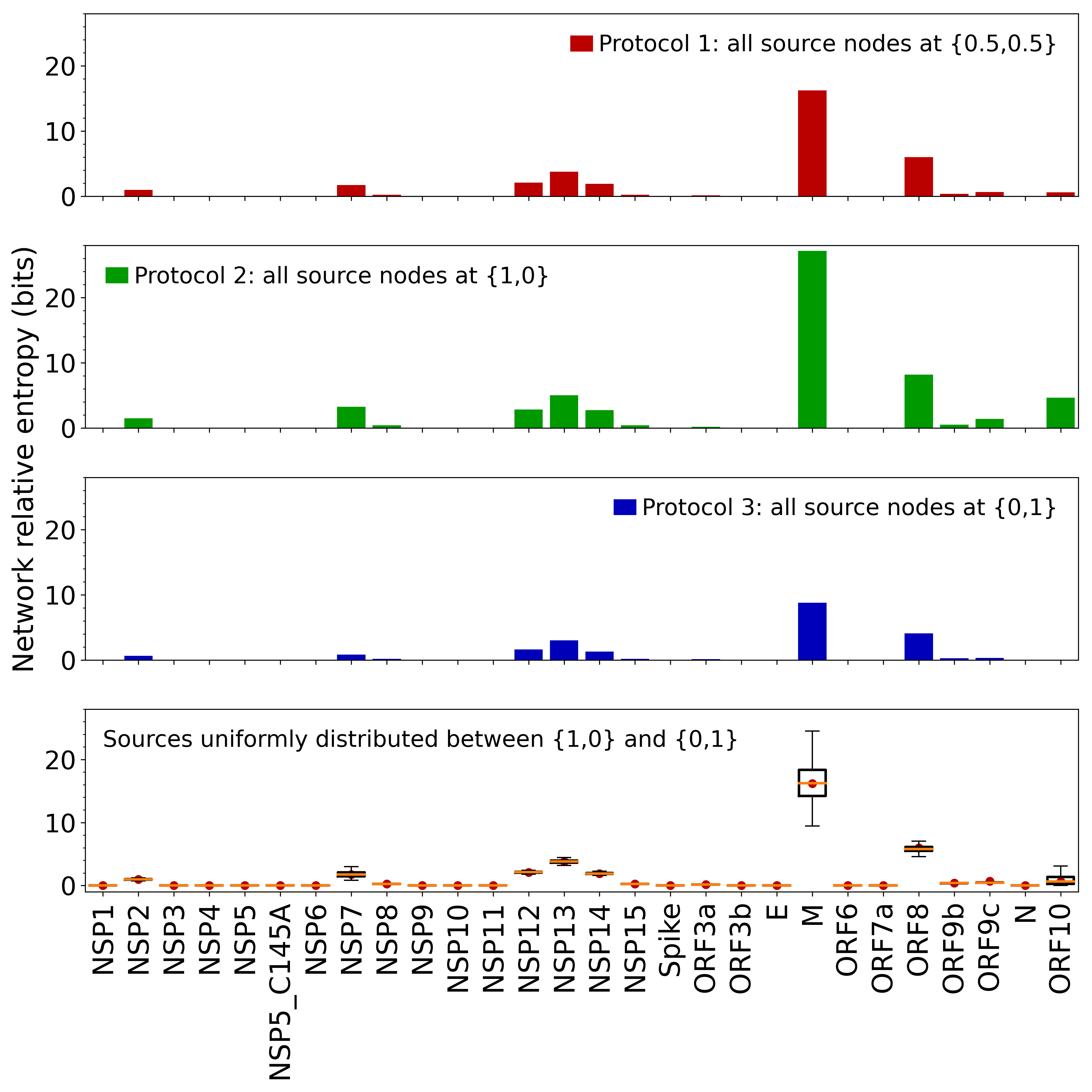} 
    }
    \end{minipage}
    \caption{\textbf{Information transfer in the immune system due to SARS-CoV2 PPIs.} (a) Communication network model of the immune system. Green nodes are proteins; blue nodes are complexes; orange nodes are molecules like ATP, ADP, or ions (Table \ref{table: node composition}). The red nodes are proteins that directly interact with the SARS-CoV-2 proteins, these nodes act as information sources in the communication network model of the immune system. The black edges are reaction like events from the Reactome database \cite{jassal2020reactome} and the green edges are protein-protein interactions from the Human Reference Interactome database \cite{luck2020reference}. (b) Network relative entropy induced by each SARS-CoV-2 protein. NSP: non structural proteins; ORF: open-reading frames for accessory factors; E: envelope protein; M: membrane protein; and N: nucleocapsid protein. In Protocol 1 all source nodes are set at the maximum entropy state, in Protocol 2 they are at maximum abundance $\{1,0\}$, and in Protocol 3 they are at minimum abundance $\{0,1\}$. The bottom-most subplot in (b) shows the distribution in the network relative entropy when each source node is uniformly distributed between maximum and minimum abundance. The distribution in network relative entropy was computed from $10^4$ samples (we compared the result against $4\times 10^4$ samples for convergence). The horizontal orange line in each box is the median, the box bounds the interquartile range (IQR), and the caps bound 1.5IQR below and above the first and the third quartiles, respectively. The red dots are the network relative entropy values from Protocol 1, which coincide with the median values.}
    \label{fig: network model}
\end{figure*}

To quantify the information transferred by SARS-CoV-2 PPIs, we first computed the reference stationary state $X_{\mathrm{ref}}$ in the absence of the SARS-CoV-2 proteins, and then constrained the SARS-CoV-2 PPI nodes as described above to compute a new stationary state $X$ using Eq. \eqref{eq: iterative info flow} (\ref{sec: PPI set}). The information transferred to a single network node $v_i$ due to the SARS-CoV-2 PPIs is given by the relative entropy (or the Kullback-Leibler divergence) in bits as \cite{cover1999elements}
\begin{equation}
\label{eq: rel entropy}
H_{\mathrm{ref}}(v_i) = \sum_{k \in  \{2i,2i+1\}} X[k] \log_2 \frac{X[k]}{X_{\mathrm{ref}}[k]} \quad .
\end{equation}
The network relative entropy $H_{\mathrm{ref}}(X)$ is the sum of the relative entropy of all the unconstrained network nodes. We computed the information transferred due to SARS-CoV-2 PPIs for multiple reference states (Fig. \ref{fig: all protocols}). In Protocol 1 all the source nodes are at the maximum entropy state $\{0.5,0.5\}$, which results in a $X_{\mathrm{ref}}$ where all the other nodes are also at maximum entropy. Then for each SARS-CoV-2 protein we constrained the corresponding subset of SARS-CoV-2 PPI nodes at $\{0,1\}$ to compute the stationary state $X$ using Eq. \eqref{eq: iterative info flow} (PPI set, \ref{table: PPI set}). In this protocol, we found that the SARS-CoV-2 proteins transfer between 0.15 bits (ORF3a protein) to 16.2 bits (M, membrane protein) of information (Fig. \ref{fig: all protocols}). Each bit of information transfer is effectively a reduction in the number of variables in the network. For example, during the interaction with SARS-CoV-2 NSP2, the network has a total of 2921 unconstrained nodes, therefore the maximum possible $H_{\mathrm{ref}}(X)$ is 2921 bits, at which the NSP2 protein completely determines the state of the entire immune network. Whereas, the actual $H_{\mathrm{ref}}(X)$ for NSP2 is 0.97 bits, or effectively it controls 1 out of the 2921 variables. A higher network relative entropy means that the particular SARS-CoV-2 protein is a more influential regulator of the immune system. When all source nodes are at high abundance $\{1,0\}$, then the SARS-CoV-2 PPIs produce a higher relative entropy (Protocol 2 in Fig. \ref{fig: all protocols}). When all source nodes are at $\{0,1\}$ then the information transferred is lower (Protocol 3 in Fig. \ref{fig: all protocols}). Or, when the effect of SARS-CoV-2 PPIs on protein function is opposite to abundances in the reference state, the information transferred is higher. The variability in $H_{\mathrm{ref}}(X)$ for the three protocols show that SARS-CoV-2 PPIs signaling is dependent on the proteomic and metabolomic profile (or abundances) of infected individuals. However, in a realistic scenario all the source nodes are unlikely to be in the same state. So, we computed the distribution of the network relative entropy using a Monte Carlo (MC) simulation. For each MC sample, we constrained each source node to a state uniformly distributed between $\{1,0\}$ and $\{0,1\}$ to first compute the reference state and then the state under SARS-CoV-2 PPIs. The median network relative entropy is equal to the value computed in Protocol 1, because unbiased interference of information drives the nodes closer to the maximum entropy state. The dispersion in information transfer, or the inter-quartile range of $H_{\mathrm{ref}}(X)$, is 4 bits for the M protein, but is around 1 bit or less for all the other SARS-CoV-2 proteins. Protocol 1 is particularly useful because it shows the information transferred only by SARS-CoV-2 PPIs, without assistance from the other source nodes in the network, and quantifies how many variables of the immune system are controlled by the viral proteins. 

\section*{\label{sec: BP analysis}Effect on biological processes}

\begin{figure*}[!ht]
    \includegraphics[width=0.99\textwidth]{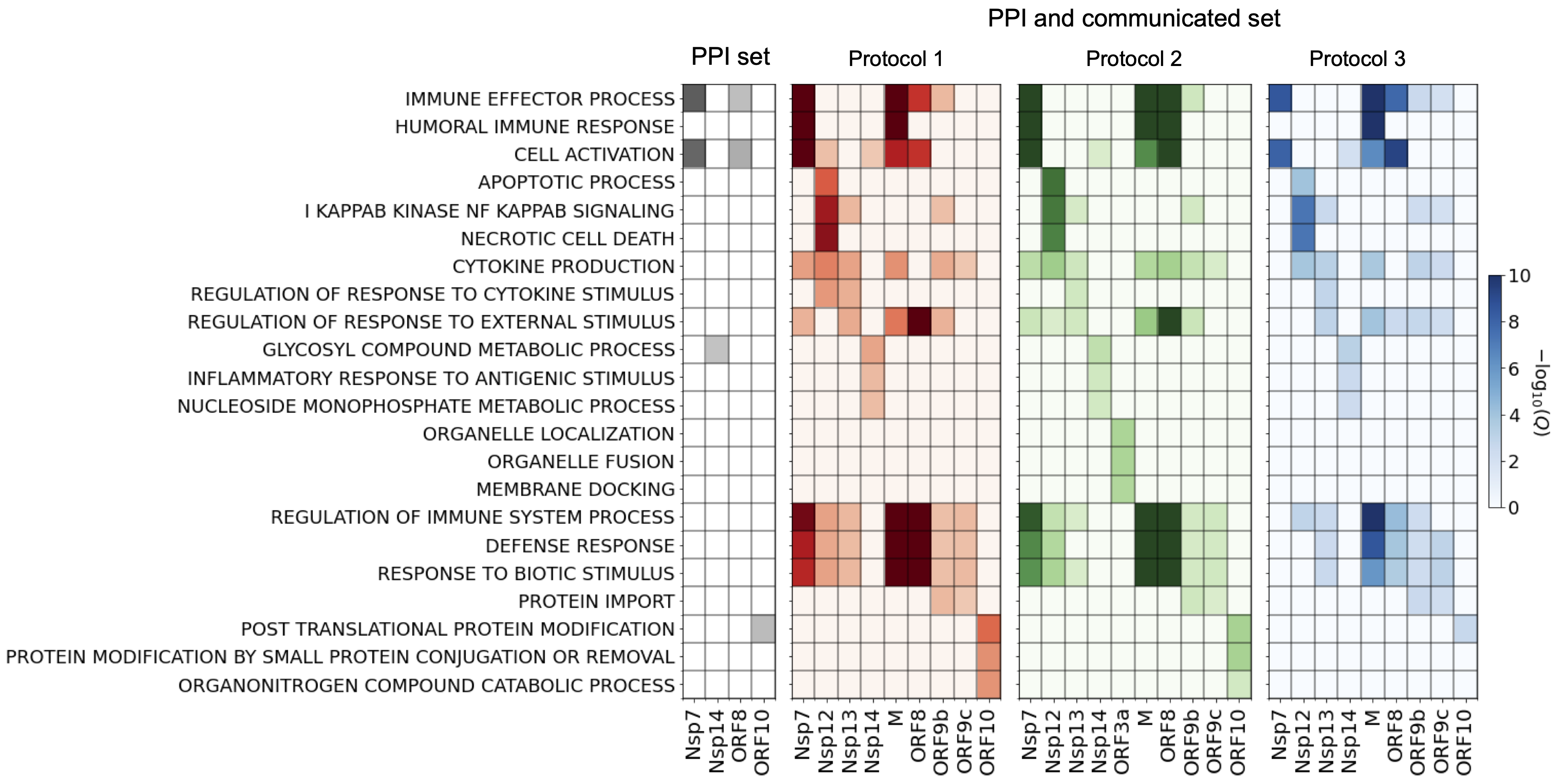} 
    \caption{\textbf{Significant biological processes due to communication from SAR-CoV-2 PPIs.} Each grid plot is the result of a Gene Ontology (GO) overrepresentation analysis for a set of proteins associated with each of the SARS-CoV-2 proteins. The enrichment analysis was done using the GOBP gene sets in the MSigDB database (\ref{sec: ORA}) \cite{subramanian2005gene,liberzon2015molecular}. The significant BPs are labeled using their adjusted $p$-value ($Q$) after multiple hypothesis testing. The first block of $Q$ values (shaded gray) is for the PPI set for each SARS-CoV-2 protein, and the remaining three blocks are for the sum of the PPI and communicated sets. We identified the significant BPs using the hypergeometric test and a false-discovery rate of 1\% \cite{thissen2002quick}, which is the same value used by the authors of the SARS-CoV-2 protein interaction map \cite{gordon2020sars}. The top 3 significant BPs for each SARS-CoV-2 protein in Protocol 2 were selected for this figure (\textit{top BPs}), because Protocol 2 produces the highest amount of information transfer in the network. We show the $Q$ values only for the top BPs using the proteins from the PPI set, Protocol 1, and Protocol 3, to compare the effect of information transfer. The complete list of all significant BPs for each protocol is in \ref{sec: ORA}.}
    \label{fig: all protocols go}
\end{figure*}

To determine the effect of communication from SARS-CoV-2 PPIs on biological processes (BPs), we first identified the immune system proteins that receive $H_{\text{ref}}(v_i) \geq 0.01$ bits of information (or the \emph{communicated set} of proteins) (\ref{sec: communicated proteins}). For each SARS-CoV-2 protein, we added the communicated set with the PPI set, and performed Gene Ontology overrepresentation analysis for BPs \cite{ashburner2000gene,storey2003positive,storey2003statistical} (Fig. \ref{fig: all protocols go}, \ref{sec: ORA}). Including the communicated set not only identifies more BPs as significant (Fig. \ref{fig: all protocols go}), but also shows that more than one SARS-CoV-2 protein can influence the same BP due to information transfer (\textit{e.g.} immune effector process and cell activation in Fig. \ref{fig: all protocols go}). The effect of COVID-19 on cytokine production and inflammation is well known \cite{hu2021cytokine,tay2020trinity}. But several of the other BPs that are significant due to communication have also been reported. These include, rapid decay in humoral immunity \cite{ibarrondo2020rapid,ward2020declining}, increased apoptosis in infected cells \cite{koupenova2021sars,li2020sars}, localization of organelles, especially endoplasmic reticulum, endosome, and lysosome \cite{zhang2020systemic,wang2020cross}, modulation of catabolic recycling processes \cite{gassen2021sars}, and protein import into the mitochondria \cite{mehrzadi2021sars}, to name a few. Additionally, dysregulation of some of the cell activation processes due to COVID-19, release of myeloid dendritic cells \cite{vabret2020immunology,mann2020longitudinal} and lymphocyte production \cite{azkur2020immune,laing2020dynamic}, have also been reported. 

\begin{figure*}[!ht]
    \begin{minipage}{0.95\textwidth}
    \subfloat[\label{fig: N-dependent ORA}]{%
    \includegraphics[width=0.95\textwidth]{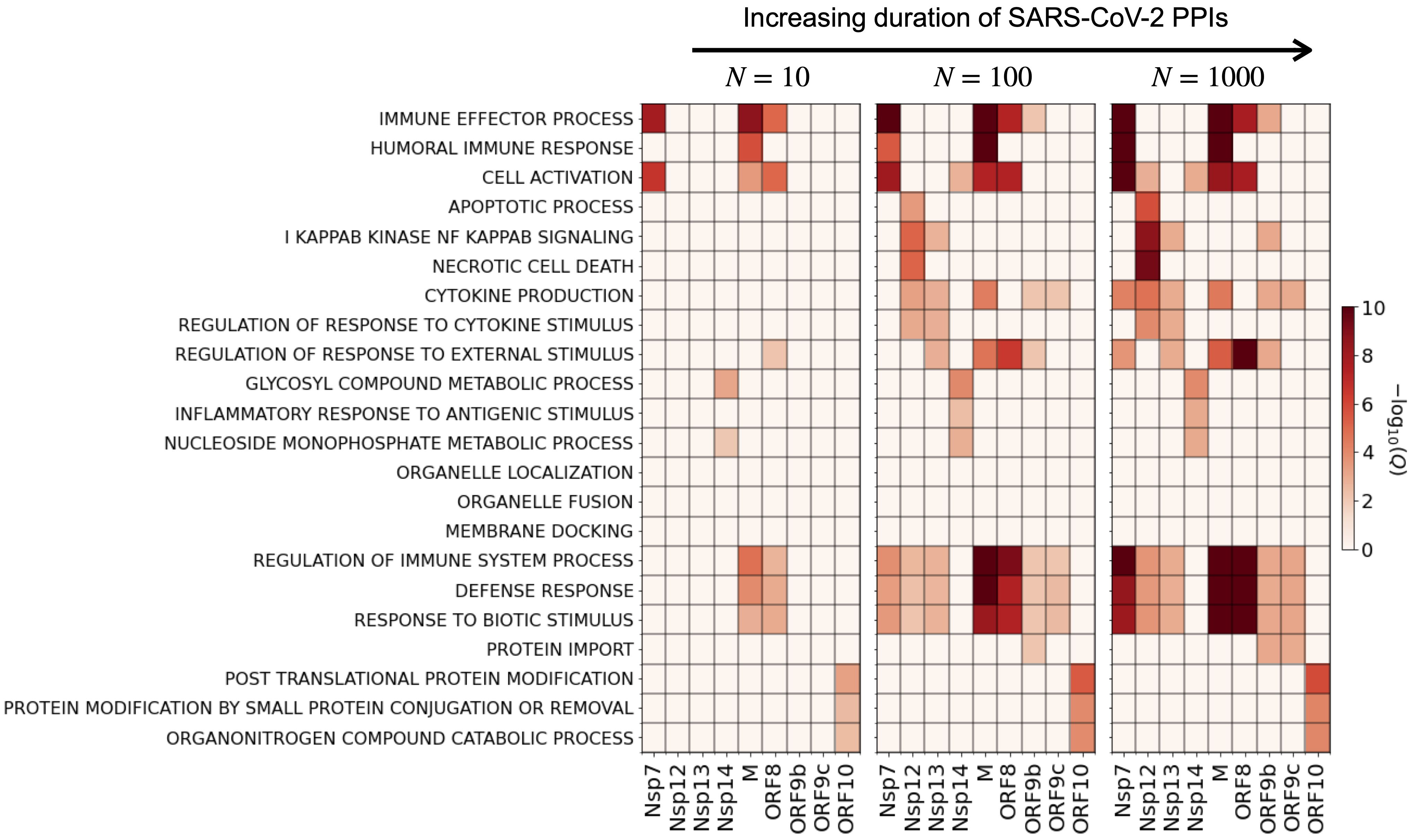}
    }
    \end{minipage} \\
    \begin{minipage}{0.98\textwidth}
    \subfloat[\label{fig: ranked BPs}]{%
    \includegraphics[width=0.5\textwidth]{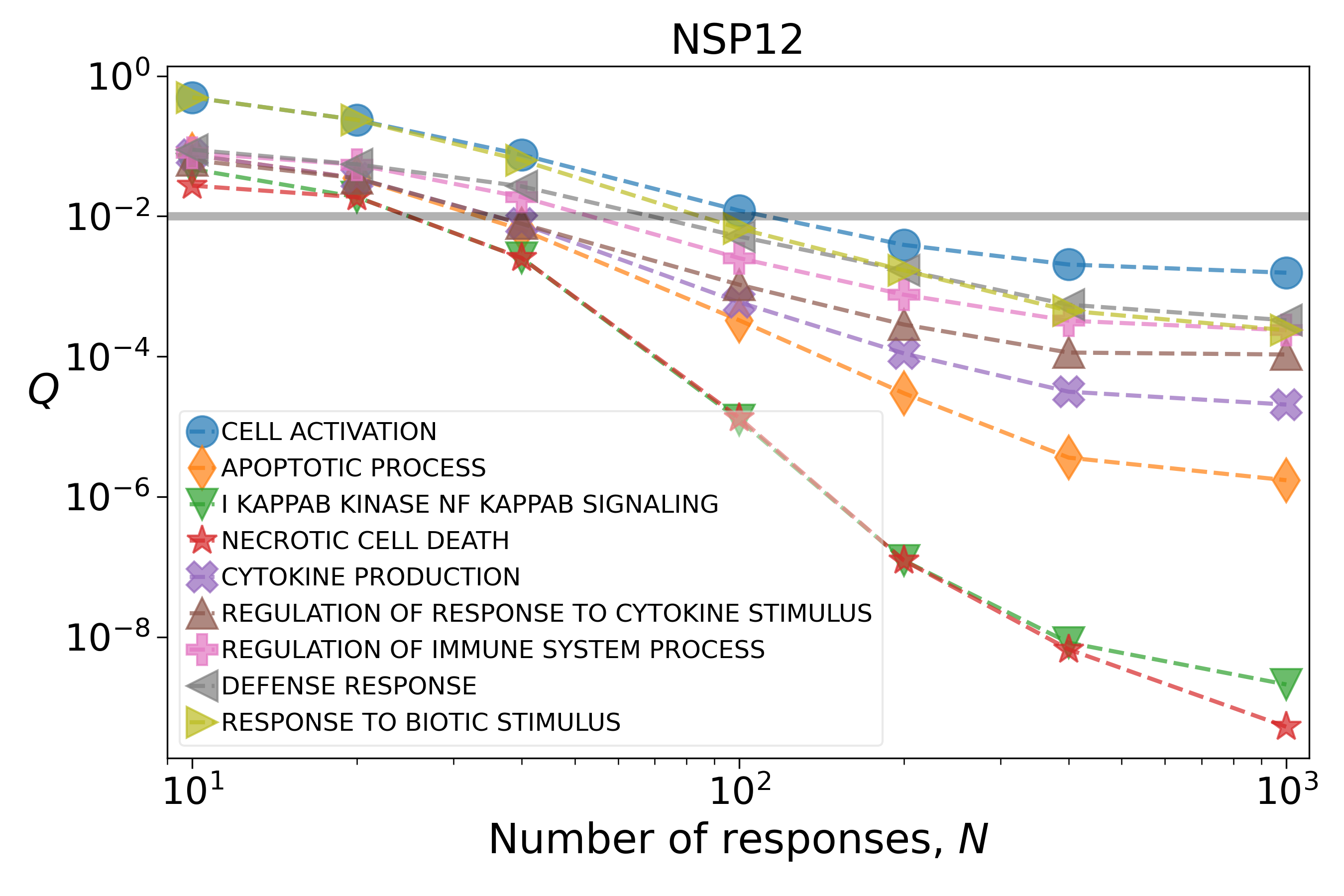} 
    \includegraphics[width=0.5\textwidth]{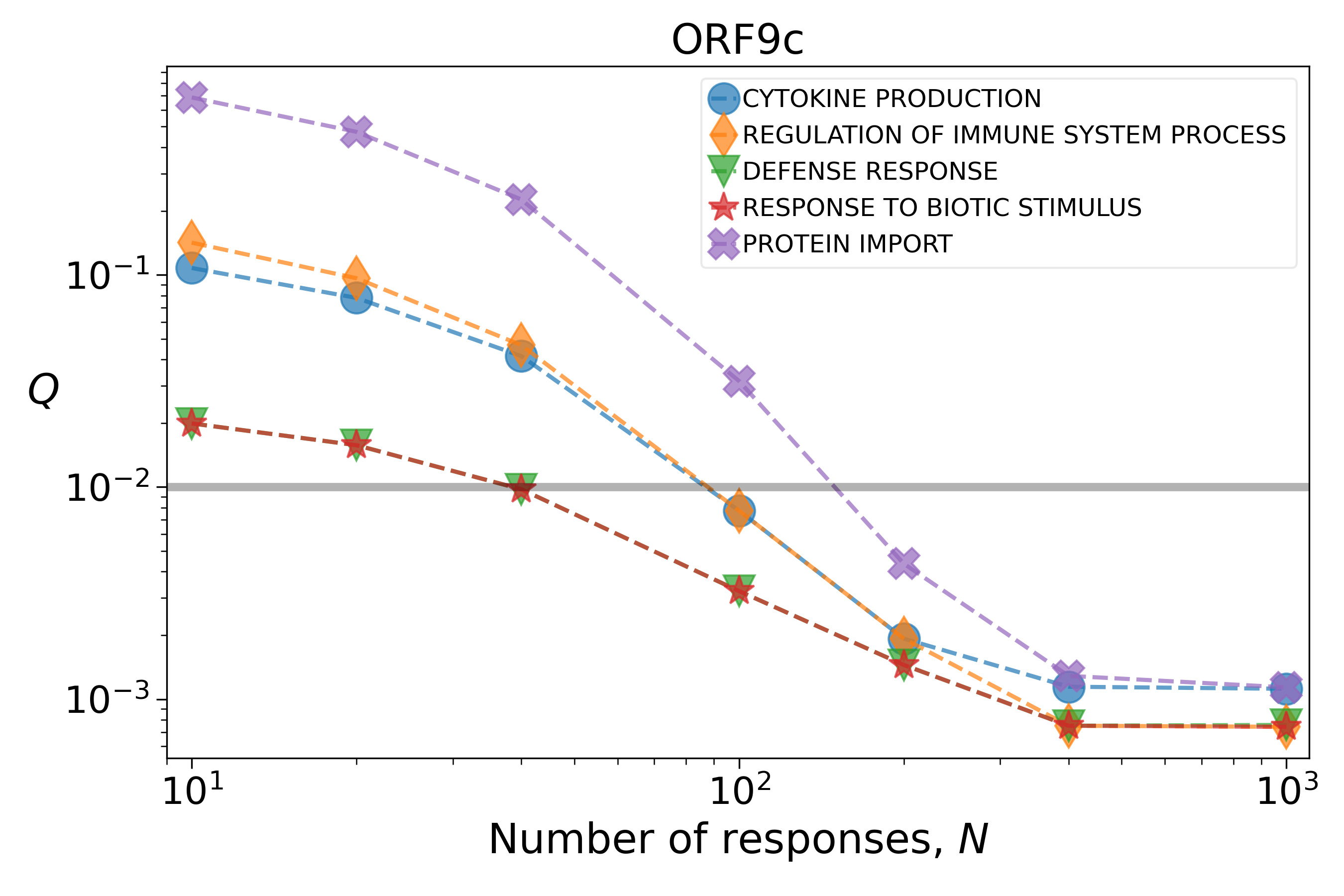} 
    }
    \end{minipage}
    \caption{\textbf{Significant biological processes as a function of the duration of SARS-CoV-2 PPIs.} (a) Each block of $Q$ values is from enrichment analysis after a finite number of responses. $N$ is directly proportional to the duration of SARS-CoV-2 PPIs, or the duration of SARS-CoV-2 PPIs $\approx N \times$ average response time of immune system reactions. The set of BPs chosen for this plot are the same as in Fig. 2. The $Q$ values were computed from a Monte Carlo simulation with $10^4$ samples (\ref{sec: en-ORA}). For each MC sample the communicated set of proteins is a random variable, which produces a set of $p$-values for the GOBP terms. The $N$-dependent $Q$ values were computed by applying the Benjamini-Hochberg procedure to the average $p$-values from the MC simulation. (b) Finite response dependent $Q$ values for the significant BPs associated with SARS-CoV-2 NSP12 and ORF9c proteins. Similar plots for other SARS-CoV-2 proteins are in \ref{sec: en-ORA}. The horizontal gray line shows the false-discovery rate used to identify the significant BPs, which is the same value used by Gordon \textit{et. al.} \cite{gordon2020sars}. Intersection of the gray line and the $Q$-$N$ curve for a BP estimates the number of responses after which the BP becomes significant for dysregulation.}
    \label{fig: effected proteins}
\end{figure*}

Using the information transferred by SARS-CoV-2 PPIs we can address the question: how might the significant BPs depend on the duration of SARS-CoV-2 PPIs? We answer this question using Protocol 1, where the reference state of a protein is $\{0.5,0.5\}$. If a protein is actually in the reference state, then after $N$ responses (or samples) of the protein abundance the empirical state can be away from $\{0.5,0.5\}$ due to stochasticity. From large deviations theory, the probability of observing an empirical state with a relative entropy greater than or equal to $H_{\mathrm{ref}}(v_i)$ after $N$ responses is $2^{-H_{\mathrm{ref}}(v_i)N}$ \cite{varadhan1984large,cover1999elements} (\ref{sec: en-ORA}). To account for this stochastic deviation from the reference state, which can occur in the absence of SARS-CoV-2 PPIs, we computed the $Q$-value of BPs as a function of the number of responses. We performed a Monte Carlo simulation where for each MC sample we eliminated proteins from the communicated set with the probability $2^{-H_{\mathrm{ref}}(v_i)N}$ and computed a set of raw $p$-values for the GOBP terms. Then we performed the multiple hypothesis correction on the averaged set of $p$-values to identify the significant BPs as a function of $N$ (\ref{sec: en-ORA}). At shorter duration ($N=10$) several BPs may not appear susceptible to dysregulation, but at longer duration ($N=1000$) the full consequence of communication from SARS-CoV-2 PPIs is revealed (Fig. \ref{fig: N-dependent ORA}). Different SARS-CoV-2 proteins can take different number of responses to effect the same BP. So, computing the $Q$-$N$ curves for each of the significant BPs is more insightful compared to only determining the significant BPs after a long duration of SARS-CoV-2 PPIs (Fig. \ref{fig: ranked BPs} for NSP12 and ORF9c, figures for the other SARS-CoV-2 proteins in \ref{sec: en-ORA}). We found that the effect of NSP12 on necrotic cell death can occur earlier than on cell activation, and the effect of ORF9c on defense response can occur earlier than on protein import. The $Q$-$N$ curves (Fig. \ref{fig: ranked BPs}) may answer clinically relevant questions, like the consequences of any delay in administering protease inhibitors during COVID-19. Protease inhibitors, like nirmatrelvir/ritonavir (sold as Paxlovid), reduce viral replication and consequently decrease the duration of SARS-CoV-2 PPIs. Therefore, using the $Q$-$N$ curves we can identify the BPs that can get dysregulated due to any delay in the use of protease inhibitors. 

\section*{\label{sec: therapies}Evaluation of candidate therapies}
\begin{figure*}[!ht]
    \begin{minipage}{\textwidth}
    \subfloat[\label{fig: drug net}]{%
    \includegraphics[height=0.4\textwidth]{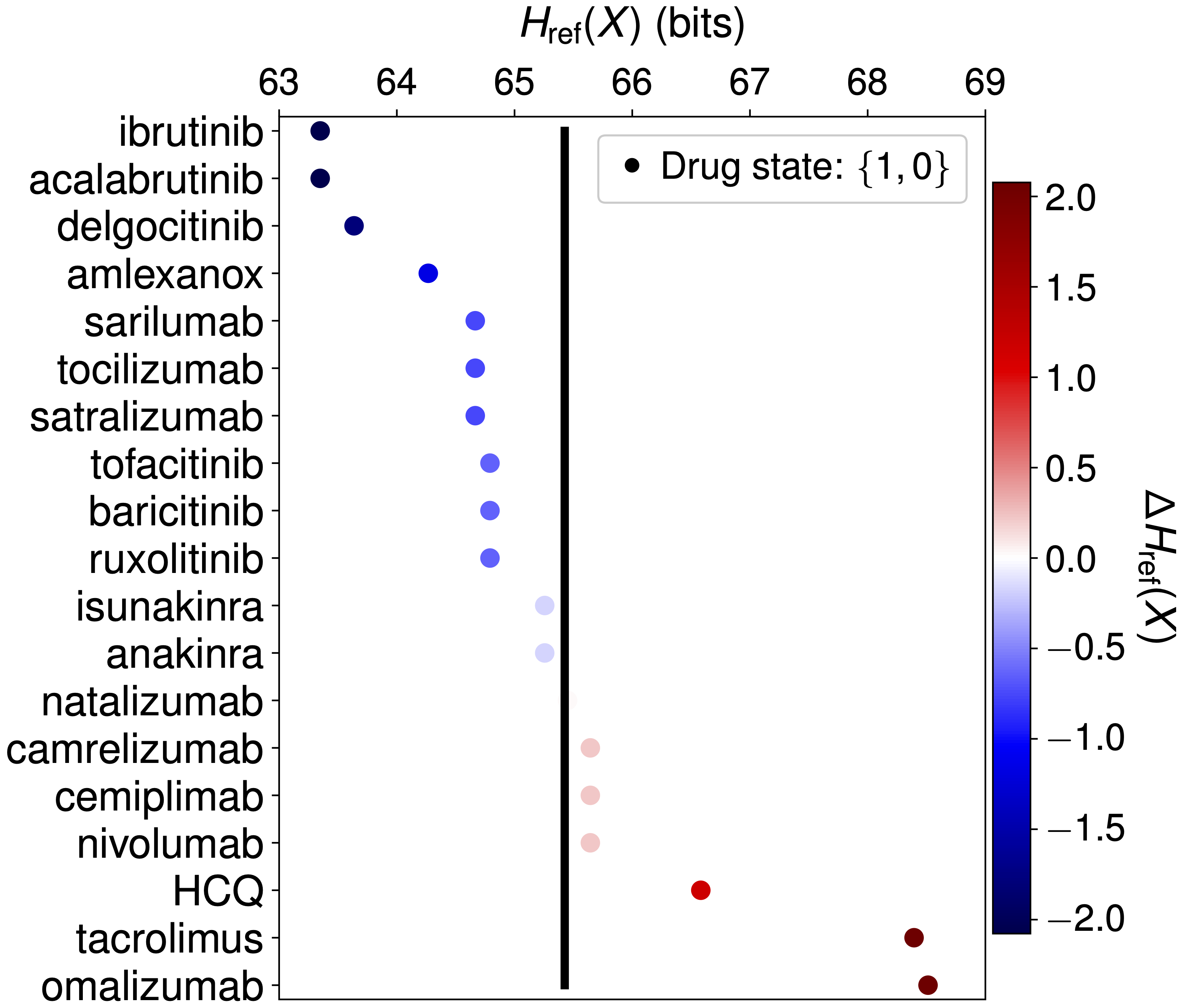} 
    }
    \subfloat[\label{fig: drug split}]{%
    \includegraphics[height=0.4\textwidth]{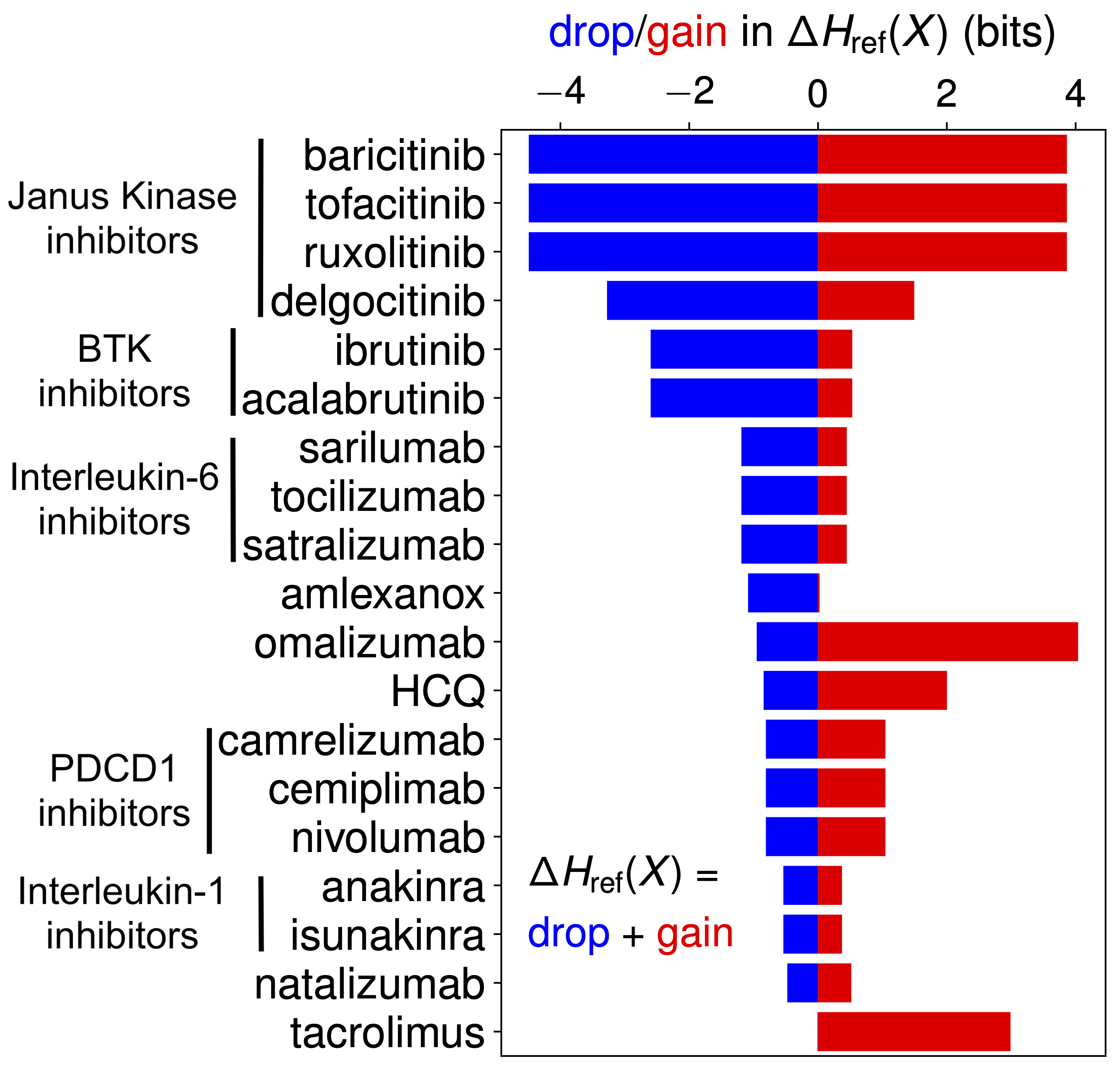}
    } \\
    \subfloat[\label{fig: protein net}]{%
    \includegraphics[height=0.375\textwidth]{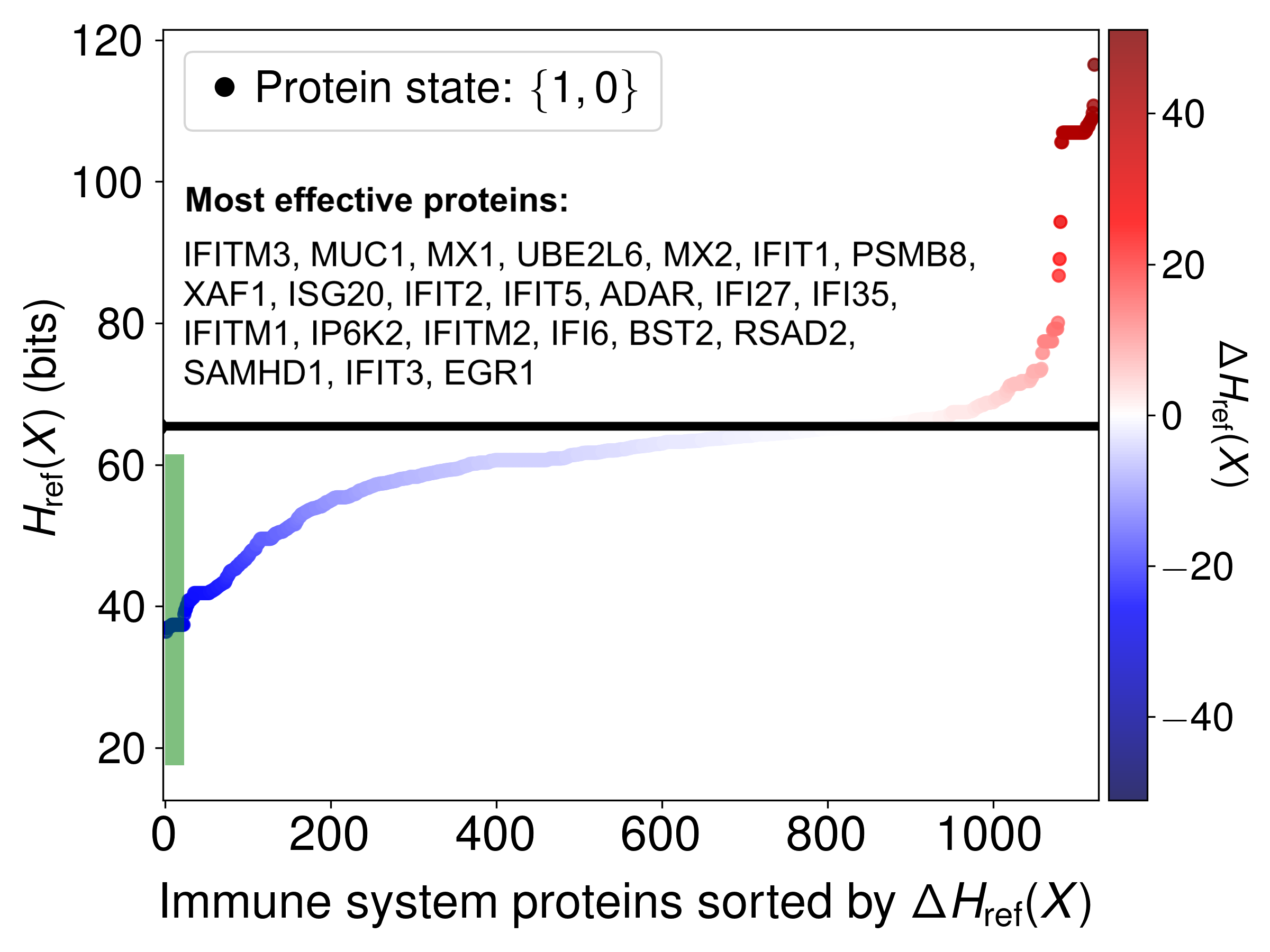}
    }
    \subfloat[\label{fig: protein split}]{%
    \includegraphics[height=0.375\textwidth]{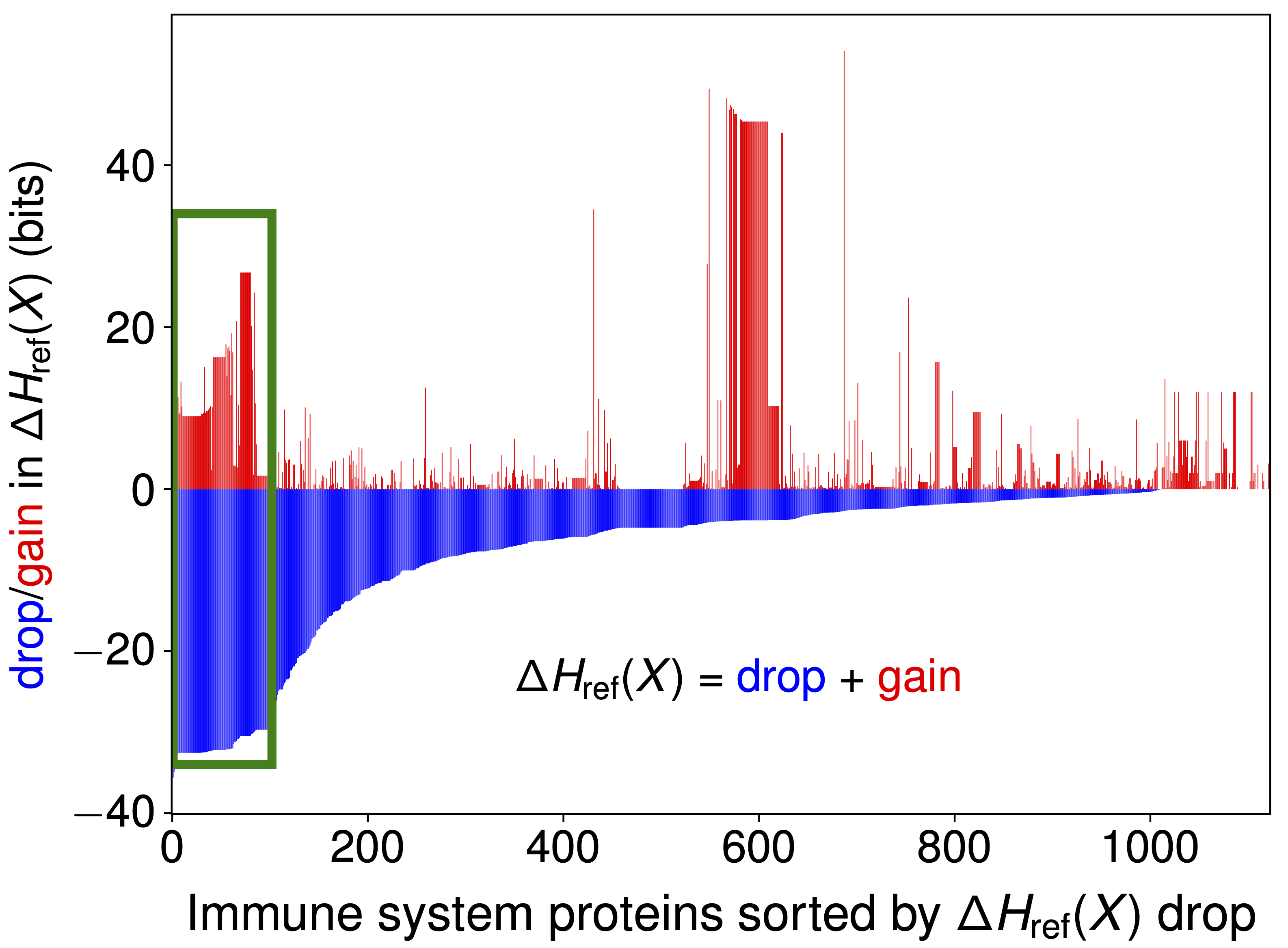}
    } 
    \end{minipage}
    \caption{\textbf{Countering the information transfer from the SARS-CoV2 PPIs using therapies.} (a) Effectiveness of chemical and protein drugs in reducing the information transfer from SARS-CoV-2 PPIs. The black vertical line is the network relative entropy when all the protein nodes in SARS-CoV-2 PPIs (\ref{table: PPI set}) are at $\{0,1\}$. The solid dots are network relative entropies when a drug is present in high abundance. The colorbar is scaled according to the net change in network relative entropy due to a drug. (b) The drop (blue) and gain (red) components of $\Delta H_{\mathrm{ref}}(X)$ due to drugs. $\Delta H_{\mathrm{ref}}(X)$ in (a) is the sum of the red and the blue bars. The list of drugs is sorted by the drop value instead of $\Delta H_{\mathrm{ref}}(X)$. Data for (a) and (b) are in SIdata5. (c) Effectiveness of artificial overexpression of immune system proteins in reducing information transfer from the SARS-CoV-2 PPIs. The horizontal line is the network relative entropy when all the protein nodes in SARS-CoV-2 PPIs (\ref{table: PPI set}) are at $\{0,1\}$. The immune system proteins, 1122 in total, are sorted in the ascending order of $\Delta H_{\mathrm{ref}}(X)$. The list of sorted proteins is in SIdata6. The green rectangle highlights the 23 proteins that reduces the information transfer from SARS-CoV-2 PPIs by 28 to 30 bits. The names of these proteins are listed in the text box. (d) The drop (blue) and gain (red) in network relative entropy due to artificial overexpression of proteins. The 1122 immune system proteins are sorted by their drop value. The green box highlights the region where the 23 most effective proteins from (c) are located. Protein ranking by the $\Delta H_{\mathrm{ref}}(X)$ drop component is in SIdata6.}
    \label{fig: counter SARS-CoV-2 PPI}
\end{figure*}

To use the communication network model for evaluating candidate therapies, we computed the reduction in the information transfer from SARS-CoV-2 PPIs due to drugs. We demonstrated this technique using the maximum entropy reference state (as in Protocol 1 in Fig. \ref{fig: all protocols}) and then applying the constraint $\{0,1\}$ on all SARS-CoV-2 PPI nodes together (all proteins in Table \ref{table: PPI set}), which produces a total network relative entropy of $65.4$ bits. The Reactome database contains 19 drugs that interact with the immune system, majority of which are immunomodulators \cite{NIHguide}. We constrained each of the drugs to the state $\{1,0\}$, one at a time, and computed the subsequent change in network relative entropy, $\Delta H_{\mathrm{ref}}(X)$ (Fig. \ref{fig: drug net}). The maximum net reduction in the SARS-CoV-2 PPIs signaling was 2 bits (for ibrutinib and acalabrutinib) and the maximum net increase was 3 bits (for omalizumab). Or, at best some of the drugs can effectively \textit{free} 2 out of the 65 variables of the immune system that are under the control of SARS-CoV-2 PPIs. 12 out of the 19 drugs produced less than $\pm 1$ bit change in the network relative entropy, indicating that majority of these 19 drugs have minimal effect in countering the signaling from SARS-CoV-2 PPIs. However, a drug can reduce the information transfer from SARS-CoV-2 PPIs in some nodes of the immune system (drop), but it can also cause additional information transfer in other nodes of the immune system (gain). So, for a more nuanced evaluation of the efficacy of a drug, we identified the network nodes that have a drop in relative entropy due to a drug and in which nodes there is a gain, to obtain the cumulative drop/gain values (Fig. \ref{fig: drug split}). Sorting by the drop component instead of $\Delta H_{\mathrm{ref}}(X)$ produces a different ranking of the drugs. We cannot neglect the gain component for new drugs, as we need to scrutinize the proteins that are receiving the gained information and the subsequent influence on BPs. However, if a drug is known to be safe, then we can neglect the gain component and use the drug with the largest drop value (Fig. \ref{fig: drug split}). Among the drugs that have large gain values, Janus Kinase (JAK) inhibitors, which inhibit cytokine activity, were considered to be safe \cite{winthrop2017emerging,hasni2021phase}, but their safety profile is being re-evaluated \cite{winthrop2022oral}. Omalizumab, an anti-Immunoglobulin E drug, has several side effects, including anaphylaxis \cite{vichyanond2011omalizumab}. JAK inhibitors have the best drop values, and they have been shown to increase survival during severe COVID-19 in clinical trials, but insignificant for reducing hospitalization \cite{chen2021jak,walz2021jak,satarker2021jak}. Bruton's tyrosine kinase (BTK) inhibitors, which impede the B-cell receptor pathway, have the second best drop values, and ibrutinib has been claimed to reduce pulmonary injury due to COVID-19 in a clinical trial \cite{treon2020btk}. Interleukin 6 (IL-6) inhibitors emerge as comparatively less effective in our analysis, which is in accordance with the results of clinical trials -- IL-6 inhibitors are significantly beneficial only when administered with glucocorticoids in clinical trials \cite{matthay20216,declercq2021effect}. But the limitations in our model is probably underestimating the efficacy of tocilizumab, which is one of the recommended treatments for critical COVID-19 \cite{klopfenstein2022tocilizumab}. As for the other drugs, the relatively low efficacy of Interleukin-1 (IL-1) inhibitors agrees with the clinical data \cite{cavalli2021right, tharaux2021effect}. Programmed cell death protein 1 (PDCD1) inhibitors, which are immune checkpoint inhibitors, are hypothesized to treat COVID-19 induced lymphopenia \cite{pezeshki2021immune}, but we did not find any study to support the claim. Omalizumab is still in a clinical trial for COVID-19 treatment \cite{omalizumabTrial}. The current NIH treatment guideline for COVID-19 recommends only baricitinib and tofacitinib among the JAK inhibitors and mostly against the BTK, IL-6, and IL-1 inhibitors, except for the cautious use of tocilizumab and no recommendation for or against anakinra \cite{NIHguide}. 

We similarly computed the effectiveness of artificially controlling the expression level of proteins to reduce the information transfer from SARS-CoV-2 PPIs. One at a time, we set the immune system proteins (except the SARS-CoV-2 PPI nodes) at high abundance $\{1,0\}$ and computed the subsequent change in the network relative entropy. We used artificial overexpression of proteins in this study because it is opposite to the effect of SARS-CoV-2 PPIs on protein function in our model. We found that 638 out of the 1122 proteins that we surveyed produced a better $\Delta H_{\mathrm{ref}}(X)$ than any of the 19 drugs (\emph{i.e}, more than 2 bits, Fig. \ref{fig: protein net}). In particular, we found 23 proteins that can reduce the signaling from SARS-CoV-2 PPIs by 28 to 30 bits, and this set contains 3 IFITM (interferon-induced transmembrane) proteins. Interestingly, artificial overexpression of IFITM proteins has been shown to block SARS-CoV-2 infection \textit{in vitro}, even though endogenous overexpression of IFITM proteins supports the infection \cite{prelli2021ifitm}. PSMB8 (Proteasome subunit beta type-8 protein, involved in antigen processing) has been reported to cause milder infection under endogenous overexpression using data from lung samples \cite{desterke2021hla}. Several of the other 23 proteins identified in this analysis (\emph{e.g.} MUC1, MX1, MX2, IFIT1, XAF1, EGR1) have been reported to be overexpressed during COVID-19, depending on the severity of the disease \cite{lu2021elevated,bizzotto2020sars,sa2020interplay,park2021gene,jiang2022insights}. We separated the drop/gain contributions to $\Delta H_{\mathrm{ref}}(X)$ for each of the proteins (Fig. \ref{fig: protein split}). The 23 proteins that produced the best $\Delta H_{\mathrm{ref}}(X)$ remain close to the top hundred even after ranking the proteins by the drop component. So, it is possible to avoid the proteins with high gain values without compromising on the reduction to SARS-CoV-2 PPIs signaling. Since controlled expression of the proteins listed in Fig. \ref{fig: protein net} can reduce the information transfer considerably better than any of the drugs, we may be able to create more effective treatments for SARS-CoV-2 PPIs compared to the ones in Fig. \ref{fig: drug split}.

\section*{Conclusion}
In summary, we developed a communication network model of the immune system to compute the information received by the various components of the system (proteins, complexes, \textit{etc.}) through multiple biochemical pathways. We demonstrated this model by computing the information transferred by SARS-CoV-2 proteins to the immune system due to their interaction with the human proteins, which reveals the ranking of the viral proteins based on their ability to control the immune system. Since the information transferred by the viral PPIs is sensitive to the reference state of the immune system, we can estimate the variability in the immunopathology of COVID-19 due to the proteomic and metabolomic profile of infected individuals. Enrichment analysis using the set of proteins that receive communication from SARS-CoV-2 PPIs leads to the potentially affected biological processes, many of which have been reported in the experimental and clinical data on COVID-19. Moreover, we can estimate the rate at which various biological processes approach dysregulation, or the $Q$-$N$ curves, as a function of the duration of SARS-CoV-2 PPIs. Perhaps the most interesting feature of the communication network model is the ability to compute the efficacy of therapies in reducing the information from viral PPIs and also the unwanted information transfer caused by the treatment, providing a new computational framework for drug selection. The top ranked drugs and proteins from our computations find support in the clinical and \textit{in vitro} data for COVID-19, in spite of the simplifying assumptions in the model.

The computational software that we have developed can construct similar communication network models for other modules in the Reactome database, \emph{e.g.}, Metabolism, Programmed Cell Death, Hemostasis, and can also combine multiple modules. The information transfer calculation is not restricted to viral PPIs only. We can use differential omics data to conceptualize reference and perturbed states, and can compute the information transfer due to various types of diseases and treatments.

\section*{Acknowledgement}
We would like to thank Mehdi Bouhaddou, Farhad Maleki, and the Reactome helpdesk team, especially Robin Haw, for helpful information during this work. We would like to thank Charles H. Camp, William Valiant, and Nancy J. Lin for their comments on the manuscript. 

\section*{Code Availability}
The source code to construct the communication network from the Reactome and the Interactome databases and to perform the subsequent computations is available at \href{https://github.com/sarkar-s/CoRe.git}{github.com/sarkar-s/core}.

\end{spacing}

\clearpage 

\bibliography{references}

\end{document}


\title{Supplementary Information: Communication network model of the immune system identifies the impact of interactions with SARS-CoV-2 proteins}

\author{Swarnavo Sarkar}
\email{swarnavo.sarkar@nist.gov}
\email{swarnavo.sarkars@gmail.com}
\affiliation{National Institute of Standards and Technology, Gaithersburg, MD 20899, USA}

\maketitle

\tableofcontents

\beginsupplement

\section{\label{sec: network construction}Construction of the immune system communication network}

To construct the communication network model of the immune system we used the Reactome biological pathways \cite{jassal2020reactome} and the Human Reference Interactome (HuRI) \cite{luck2020reference} databases. The Reactome database contains the list of biochemical reactions within the top level pathway of the Immune System. In this database, the immune system contains 1622 biochemical reactions and 176 regulatory reactions. Each reaction has a set of inputs and outputs, and each input (or output) is a PhysicalEntity in the Reactome database. Subclasses of PhysicalEntity are EntitiesWithAccessionedSequence (for proteins), Complex (for complexes), SimpleEntity (for molecules like ATP and ions like \ce{Ca^2+}), and some other classes with a smaller number of physical entities. We created the immune system communication network $G=(V,E)$, by adding the edges (or connections) between the PhysicalEntities using the input-output relationships from the ReactionLikeEvent class in the Reactome database. The ReactionLikeEvent class contains the subclasses of Reaction, BlackBoxEvent, Polymerisation, and Depolymerisation. Edges from the Reaction subclass are bidirectional, and the edges from the remaining subclasses are unidirectional. Information (or stochasticity) from each input can propagate to the information in all the outputs of an event. Therefore, for a single event, each pair of input and output is represented by an edge (or an information channel). Hence, a reaction like, \ce{A + B -> C + D +E}, contributes 6 edges to the immune system communication network, $\{\ce{A} \to \ce{C}, \ce{A} \to \ce{D}, \ce{A} \to \ce{E}, \ce{B} \to \ce{C}, \ce{B} \to \ce{D}, \ce{B} \to \ce{E}\}$. Inhibitory and negative regulation reactions reduce the abundance of the reaction output with increasing abundance of the input (or the regulator), we represent these edges as $\ce{A} \myarrow \ce{B}$. All other types of reactions, where the input increases the abundance of the output, are represented as $\ce{A} \to \ce{B}$. After constructing the network from the Reactome database, we used the HuRI database \cite{luck2020reference} to identify the pairs of proteins that interact with each other. Each pairwise interaction was added as a bidirectional edge (type $\to$) to the network (currently, the HuRI database does not contain the sign of interactions). We used the NetworkX library in Python to create the network object. Once the network was created we removed all isolates (\textit{i.e.}, nodes that are not connected to any other nodes) from the network. The total number of nodes in the network is $|V|=3011$, and the breakdown of the number of nodes by their class is in Table \ref{table: node composition}. The list of nodes and edges for the immune system communication network are in the files SIdata1 and SIdata2, respectively. 

\begin{table}[h!]
\centering
\begin{tabular}{ l|l } 
\hline
Node class & Number \\ 
\hline
EntityWithAccessionedSequence & 1141 \\ 
Complex & 1609 \\
SimpleEntity & 164 \\
OtherEntity & 46 \\
Polymer & 21 \\
GenomeEncodedEntity & 30 \\
Protein Drug & 10 \\
Chemical Drug & 9 \\
\hline
\end{tabular}
\caption{Composition of nodes in the immune system communication network. The Node class column lists the schemaClass attribute for a PhysicalEntity in the Reactome database. Protein and chemical drug nodes were removed from the network except for when computing the drop in the information from viral PPIs due to drugs. EntityWithAccessionedSequence is a derived class of GenomeEncodedEntity, so the GenomeEncodedEntity number in the table only counts the entities that are not part of the EntityWithAccessionedSequence class already.}
\label{table: node composition}
\end{table}

\section{\label{sec: channel model}Information transfer through the network}

To simulate the transfer of information in the immune system communication network, we modeled each edge of the network as a binary information channel. Reactions where the abundance of the input increases the abundance of the output (including positive regulation reactions) were modeled as $v_i \to v_j$, with the input and the output state related as
\begin{equation}
\label{eq: up channel}
\begin{Bmatrix}
P(v_j=1) \\
P(v_j=0)
\end{Bmatrix}
=
\underbrace{\begin{bmatrix}
1-\rho & \rho \\
\rho & 1-\rho
\end{bmatrix}}_{T_{\mathrm{up}}(\rho)}
\begin{Bmatrix}
P(v_i=1) \\
P(v_i=0)
\end{Bmatrix}
\end{equation}
where $\rho<0.5$ is the channel error that reduces information transfer from the input $v_i$ to the output $v_j$. Reactions where the abundance of the input decreases the abundance of the output (inhibitory and negative regulation reactions) were modeled as bit flip information channels, represented as $v_i \myarrow v_j$, with the input and the output state related as 
\begin{equation}
\label{eq: down channel}
\begin{Bmatrix}
P(v_j=1) \\
P(v_j=0)
\end{Bmatrix}
=
\underbrace{\begin{bmatrix}
\rho & 1-\rho \\
1-\rho & \rho
\end{bmatrix}}_{T_{\mathrm{down}(\rho)}}
\begin{Bmatrix}
P(v_i=1) \\
P(v_i=0)
\end{Bmatrix} \quad .
\end{equation}
The matrices in equations \eqref{eq: up channel} and \eqref{eq: down channel} are transition matrices for the information channel model for individual edges in the network. We built the global transition matrix for the entire immune system network using the method described in \cite{sarkar2021information}. Let the adjacency matrix of the graph be $A$, which is a $|V|$-by-$|V|$ matrix, with $A[i,j] = 1$ if there is an edge from $v_i$ to $v_j$, else $A[i,j]=0$. The global transition matrix $C_G$ is a $2|V|$-by-$2|V|$ dimensional matrix such that
\begin{equation}
\label{eq: global transition}
C_G[2j,2j+1;2i,2i+1] = 
\begin{cases}
\frac{w_{ij}}{\text{deg}^{-}(v_j)} T_{\mathrm{up}}(\rho) \quad & \text{if } A[i,j]=1  \text{ and } v_i \to v_j\\
\frac{w_{ij}}{\text{deg}^{-}(v_j)} T_{\mathrm{down}}(\rho) \quad & \text{if } A[i,j]=1  \text{ and } v_i \myarrow v_j\\
0_{2,2} \quad & \text{if } A[i,j]=0 \\
I_2 \quad & \text{if } \text{deg}^{-}(v_j)=0 \text{ and } i=j
\end{cases}
\end{equation}
where $\text{deg}^{-}(v_j)$ is the in-degree of node $v_j$, or the node that is receiving information from the source $v_i$. The weights $w_{ij}$ determine the type of interference of information. For unbiased interference all weights are $w_{ij}=1$. For biased interference the weights has to satisfy the conditions $0<w_{ij}$ and $\sum_j w_{ij} = \text{deg}^{-}(v_j)$. We used unbiased interference in our calculations. Proteomic and metabolomic profile data of human cells \cite{kim2014draft} can be used to create a data-driven biased interference model. We also used $\rho=0$ for the results presented in the main text. So, our results capture the information loss due to signal interference rather than communication noise. Non-zero communication error will further decrease the network relative entropy from the SARS-CoV-2 PPIs, communicate to a smaller set of proteins, and thereby affect a smaller set of biological processes. Data on the rate constants of human proteins and transcripts \cite{friedel2009conserved,cambridge2011systems,hausser2019central} can be used to estimate the $\rho$ value for each reaction \cite{sarkar2022nearly}. 

We represent the global state of the communication network using a column vector $X$ with the dimension $2|V|$. As mentioned in the main text, the $i^{th}$ 2D block of $X$ contains the state of the node $v_i$, $\{P(v_i=1),P(v_i=0)\}$. The transfer of information through the network is computed using the equation
\begin{equation}
\label{eq: global communication}
X^{(k+1)} = C_G X^{(k)}
\end{equation}
where each multiplication by $C_G$ causes information to transfer from a network node to its nearest neighbors. The stationary state of the network is obtained when $\|X^{(k+1)} - X^{(k)}\|_1 < \epsilon$, where $\|\cdot\|_1$ is the L1-norm and the $\epsilon$ is an error tolerance, we used $\epsilon = 10^{-4}$ in all of our calculations, which we found sufficient to achieve convergence in the network relative entropy for the 3 protocols. Finally, it is not necessary to model each node as a binary variable and each edge as a 1 bit information channel. We can easily construct the global transition matrix $C_G$ using a $n>1$ bit channel model for each edge, using the procedure described through equations \eqref{eq: up channel}, \eqref{eq: down channel}, and \eqref{eq: global transition}. In general, the state vector $X$ has a dimension $2^n|V|$ and the global transition matrix is $2^n|V|$-by-$2^n|V|$, as a function of the bit size for the channel model for each edge. 

\section{\label{sec: PPI set}Information transfer due to SARS-CoV-2 PPIs}

To compute the information transfer in the immune system network due to SARS-CoV-2 PPIs, we first identified the immune system proteins that directly interact with each SARS-CoV-2 protein using the protein interaction map published by Gordon \emph{et. al.} \cite{gordon2020sars}. The set of immune system proteins that directly interact with each of the SARS-CoV-2 proteins are in Table. \ref{table: PPI set}.

\begin{figure*}[!ht]
    \includegraphics[width=0.65\textwidth]{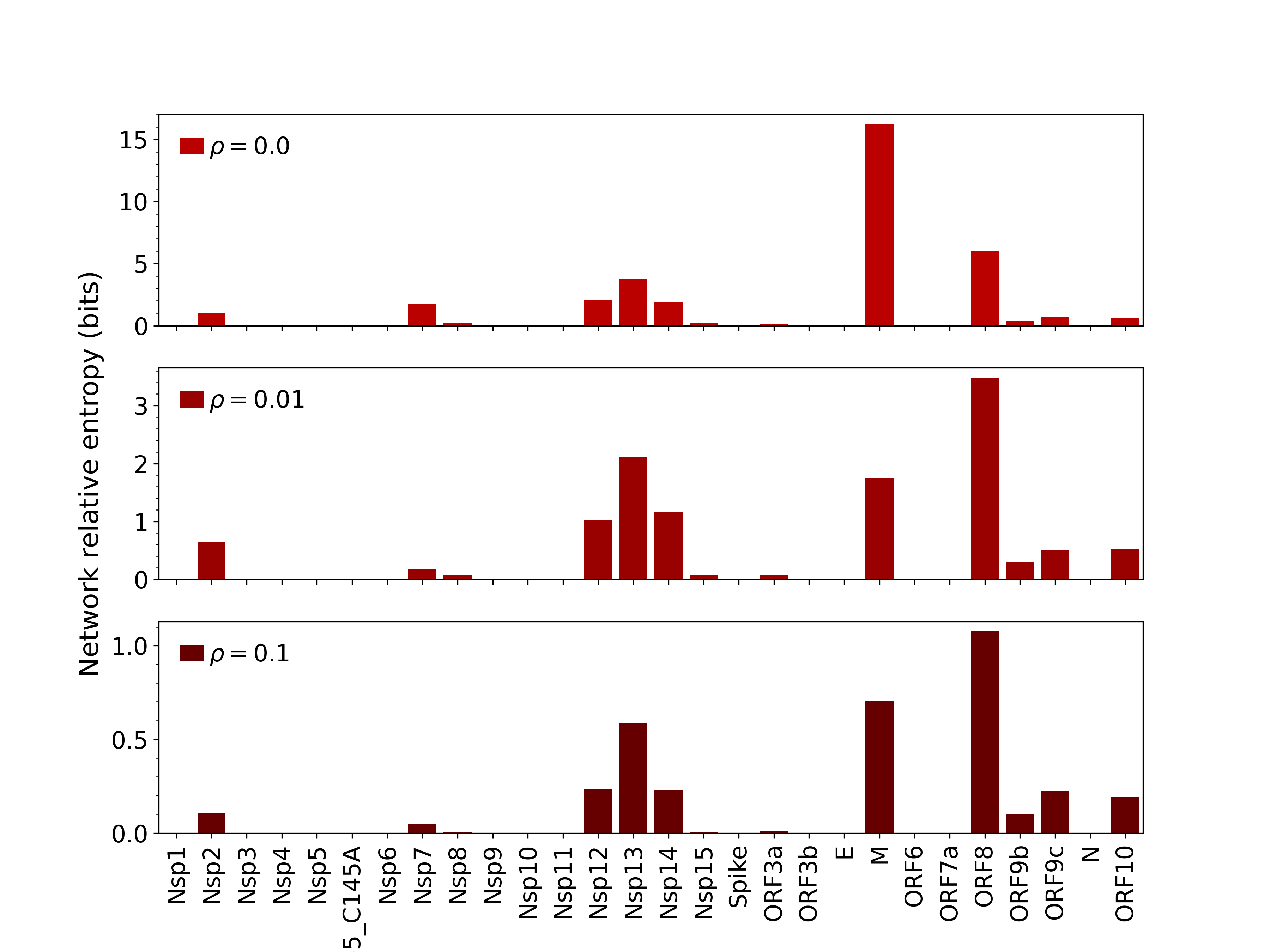} 
    \caption{Change in network relative entropy due to communication errors. The network relative entropy drops for all proteins but most prominently for the membrane (M) protein.}
    \label{fig: errors}
\end{figure*}

\begin{table}[h!]
\sffamily
\footnotesize
\centering
\begin{tabular}{ l|l } 
\hline
SARS-CoV-2 proteins & PPI set of proteins\\ 
\hline
NSP2 & SLC27A2, EIF4E2 \\ 
NSP7 & CYB5R3, RALA, RAB5C, RAB7A, RAB10, RAB14, RHOA, PTGES2, RAB18 \\
NSP8 & HECTD1 \\
NSP10 & AP2A2 \\
NSP12 & RIPK1 \\
NSP13 & RDX, ERC1, TBK1 \\
NSP14 & GLA, IMPDH2 \\
NSP15 & RNF41 \\
Spike & GOLGA7 \\
ORF3a & HMOX1 \\
E & SLC44A2 \\
M & STOM, ANO6 \\
ORF8 & ITGB1, PVR, NPC2, GGH, IL17RA, NEU1, ERP44 \\
ORF9b & TOMM70 \\
ORF9c & NLRX1, ECSIT \\
N & CSNK2B \\
ORF10 & ELOC, ELOB \\
\hline
\end{tabular}
\caption{Immune system proteins that directly interact with each of the SARS-CoV-2 proteins \cite{gordon2020sars}.}
\label{table: PPI set}
\end{table}
To determine the information transfer due to each SARS-CoV-2 protein, we constrained the corresponding set of immune system proteins from Table \ref{table: PPI set} to the state $\{0,1\}$ and computed the network relative entropy of the stationary state with respect to the reference state for each protocol. We used $\rho=0$ for the results shown in the main text. The network relative entropy values for the 3 protocols shown in Fig. 1(b) are in the file SIdata3. We present the network relative entropy values for a set of other $\rho$ values in Fig. \ref{fig: errors}. A higher communication error reduces the information transfer, which means a smaller number of immune system proteins will receive information about the SARS-CoV-2 PPIs. Network relative entropy always decreases with increasing $\rho$. But $\rho>0$ can also change the ranking of the SARS-CoV-2 proteins based on the relative entropy induced by them. 

\begin{figure*}[!ht]
    \includegraphics[width=0.65\textwidth]{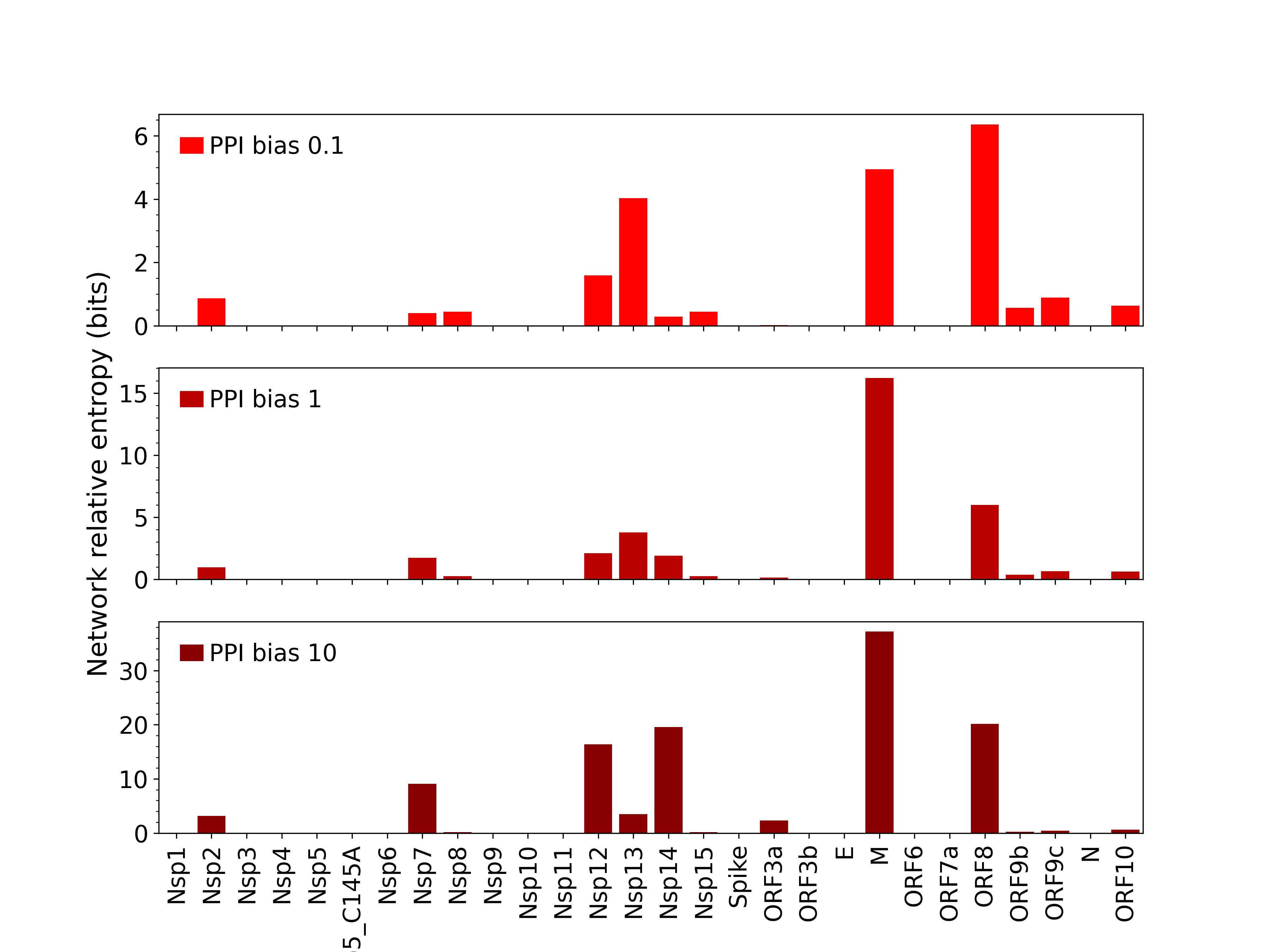} 
    \caption{Change in network relative entropy due to biased interference of information. The label PPI bias refers to the relative weights of the PPI edges compared to other edges in the immune system network.}
    \label{fig: 3 ppis}
\end{figure*}

Similarly, we demonstrate the effect of biased interference in Fig. \ref{fig: 3 ppis}. We considered three cases. In the first one we set the PPI edges in the network with weights 1/10$^{th}$ of the weights for all the other edges, so the interactions between the immune system proteins transfer less information compared to the reactions in the network. The second case is unbiased interference, which is same as the Protocol 1 results in the main text. In the third one the PPI edges have 10 times the weight compared to the other edges in the network, or protein-protein interactions are the dominant channels for communication in this case. Fig. \ref{fig: 3 ppis} shows the sensitivity of the network relative entropy to the bias in the interference of information.

\clearpage

\section{\label{sec: communicated proteins}Proteins receiving information from SARS-CoV-2 PPIs}

To identify the proteins that are receiving communication from SARS-CoV-2 PPIs, we determined the protein nodes in the communication network model of the immune system that have a relative entropy $H_{\mathrm{ref}}(v_i) \geq 0.01$ bits. Tables \ref{table: protocol 1 proteins}-\ref{table: protocol 3 proteins} list the immune system proteins that receive information from each SARS-CoV-2 protein for each of the 3 protocols. The relative entropy values for each protein are in the file SIdata4.

\begin{table}[h!]
\sffamily
\small
\centering
\begin{tabular}{ l|l } 
\hline
SARS-CoV-2 proteins & Communicated set of proteins \\ 
\hline
NSP2 & CARD9 \\ 
NSP7 & C3, C3AR1, CFB, C2, C4B, C4A, C5, C5AR1, STOM, CD55, CR1, HMOX2, CFI, CFH, CFHR3, CD46, \\
& CR2, C5AR2, PROS1 \\
NSP12 & TICAM1, RIPK3, FADD, CASP8, SARM1, CASP10 \\
NSP13 & IKBKE, TOMM70 \\
NSP14 & IL36RN, IL1F10, IL1RL2, IL1RAPL1, IMPDH1 \\
ORF3a & STX1A \\
M & C3, C3AR1, CFB, C2, C4B, C4A, C5, C5AR1, IL7R, PDCD1LG2, MCEMP1, ATP6V0C, CD55, CR1, \\ 
& CYB5R3, HMOX2, CFI, CFH, CFHR3, CD46, CR2, C6, C5AR2, PROS1, EDAR, EDA, EDA2R, KLRC1, \\
& KLRD1, KLRG1, CDH1 \\
ORF8 & IL11, IL11RA, CD47, ITGB2, ITGAM, ITGAX, OLFM4, CTSA, CD68, MUC12, MUC17, MUCL1, \\ 
& MUC5B, MUC3A, MUC1, MUC15, MUC20, MUC7, MUC4, MUC16, MUC2, MUC19, MUC3B, MUC21, \\
& MUC13, MUC6, MUC5AC, TNFRSF25, TNFSF15, TNFRSF6B, TNFSF14, TNFSF6, FCER2, NECTIN2, \\
& CRTAM, CD226, CD96, KIR2DL2, KIR2DL3 \\
ORF9b & MAVS, NLRX1 \\
ORF9c & MAP3K1, MAVS, TOMM70 \\
ORF10 & RNF7, CUL5 \\
\hline
\end{tabular}
\caption{Communicated set of proteins under Protocol 1. List of immune system proteins that receive more than or equal to 0.01 bits of information due to the SARS-CoV-2 PPIs under Protocol 1.}
\label{table: protocol 1 proteins}
\end{table}

\begin{table}[h!]
\sffamily
\small
\centering
\begin{tabular}{ l|l } 
\hline
SARS-CoV-2 proteins & Communicated set of proteins\\ 
\hline
NSP2 & CARD9 \\ 
NSP7 & C3, C3AR1, CFB, C2, C4B, C4A, C5, C5AR1, STOM, CD55, CR1, HMOX2, CFI, CFH, CFHR3, \\
& CD46, CR2, C5AR2, PROS1 \\
NSP12 & TICAM1, RIPK3, FADD, CASP8, TRAF3, SARM1, CASP10, CASP4, CASP9, CASP2 \\
NSP13 & IKBKE, SIKE1, TOMM70 \\
NSP14 & IL36RN, IL1F10, IL1RL2, IL1RAPL1, IMPDH1 \\
ORF3a & STX1A, STX3, VAMP3 \\
M & C3, C3AR1, CFB, C2, C4B, C4A, C5, C5AR1, IL7R, PDCD1LG2, MCEMP1, ATP6V0C, CD55, \\
& CR1, CYB5R3, HMOX2, CFI, CFH, CFHR3, CD46, CR2, C6, C5AR2, PROS1, EDAR, EDA, EDA2R, \\
& KLRC1, KLRD1, KLRG1, CDH1 \\
ORF8 & C3, C3AR1, CFB, C2, C4B, C4A, APP, IL11, IL11RA, CD47, ITGB2, ITGAM, ITGAX, CD55, \\
& CR1, OLFM4, CTSA, CD68, MUC12, MUC17, MUCL1, MUC5B, MUC3A, MUC1, MUC15, MUC20, MUC7, \\
& MUC4, MUC16, MUC2, MUC19, MUC3B, MUC21, MUC13, MUC6, MUC5AC, CFI, CFH, CFHR3, CD46, \\
& CR2, PROS1, TNFRSF25, TNFSF15, TNFRSF6B, TNFSF14, TNFSF6, FCER2, NECTIN2, CRTAM, \\ 
& CD226, CD96, KIR2DL2, KIR2DL3, BTN2A2, BTNL2 \\
ORF9b & MAVS, NLRX1 \\
ORF9c & MAP3K1, MAVS, TOMM70 \\
ORF10 & SOCS1, RNF7, CUL5 \\
\hline
\end{tabular}
\caption{Communicated set of proteins under Protocol 2. List of immune system proteins that receive more than or equal to 0.01 bits of information due to the SARS-CoV-2 PPIs under Protocol 2.}
\label{table: protocol 2 proteins}
\end{table}

\begin{table}[h!]
\sffamily
\small
\centering
\begin{tabular}{ l|l } 
\hline
SARS-CoV-2 proteins & Communicated set of proteins\\ 
\hline
NSP7 & STOM, HMOX2 \\
NSP12 & RIPK3, FADD, CASP8, CASP10 \\
NSP13 & IKBKE, TOMM70 \\
NSP14 & IL36RN, IL1F10, IL1RL2, IL1RAPL1, IMPDH1 \\
ORF3a & STX1A \\
M & C3, C3AR1, CFB, C2, C4B, C4A, C5, C5AR1, MCEMP1, ATP6V0C, CD55, CR1, CYB5R3, \\
& HMOX2, CFI, CFH, CFHR3, CD46, CR2, C5AR2, PROS1, EDA, EDA2R \\
ORF8 & IL11, IL11RA, CD47, ITGB2, ITGAM, ITGAX, OLFM4, MUC1, TNFRSF25, TNFSF15, \\
& TNFRSF6B, TNFSF14, TNFSF6, NECTIN2, CRTAM, CD226, CD96 \\
ORF9b & MAVS, NLRX1 \\
ORF9c & MAVS, TOMM70 \\
\hline
\end{tabular}
\caption{Communicated set of proteins under Protocol 3. List of immune system proteins that receive more than or equal to 0.01 bits of information due to the SARS-CoV-2 PPIs under Protocol 3.}
\label{table: protocol 3 proteins}
\end{table}

\section{\label{sec: ORA}Significant biological processes due to the communicated set of proteins}

To identify the biological processes that are candidates for dysregulation, we performed Gene Ontology (GO) overrepresentation analysis using the Biological Processes (BP) gene sets \cite{ashburner2000gene}. We used the GOBP gene sets in the c5 collection of the MSigDB database \cite{subramanian2005gene}. We determined the total interaction set for each SARS-CoV-2 protein, \textit{i.e.}, sum of the PPI set (\ref{table: PPI set}) and one of the communicated sets (\ref{table: protocol 1 proteins}-\ref{table: protocol 3 proteins}) depending on the protocol. Then we determined if the total interaction set is overrepresented in any of the GOBP gene sets. The $p$-value for each GOBP term was computed using the hypergeometric survival function from scipy.stats in Python. A false discovery rate (FDR) of 1\% was used for multiple hypothesis testing using the Benjamini-Hochberg correction, which provided the adjusted $p$-value ($Q$). This computation was done using the fdrcorrection function in the statsmodels library. After determining the significant GOBP terms (for FDR 1\%), we further reduced the set of significant terms by eliminating the redundant terms. To eliminate the redundant gene sets, we first surveyed all the MSigDB GOBP gene sets to identify which GOBP terms are contained in others, or which GOBP gene sets are subsets of other GOBP gene sets. After determining all the significant GOBP terms from the overrepresentation analysis, we removed the GOBP terms that are contained in other GOBP terms.

\begin{figure*}[!ht]
    \includegraphics[height=\textheight]{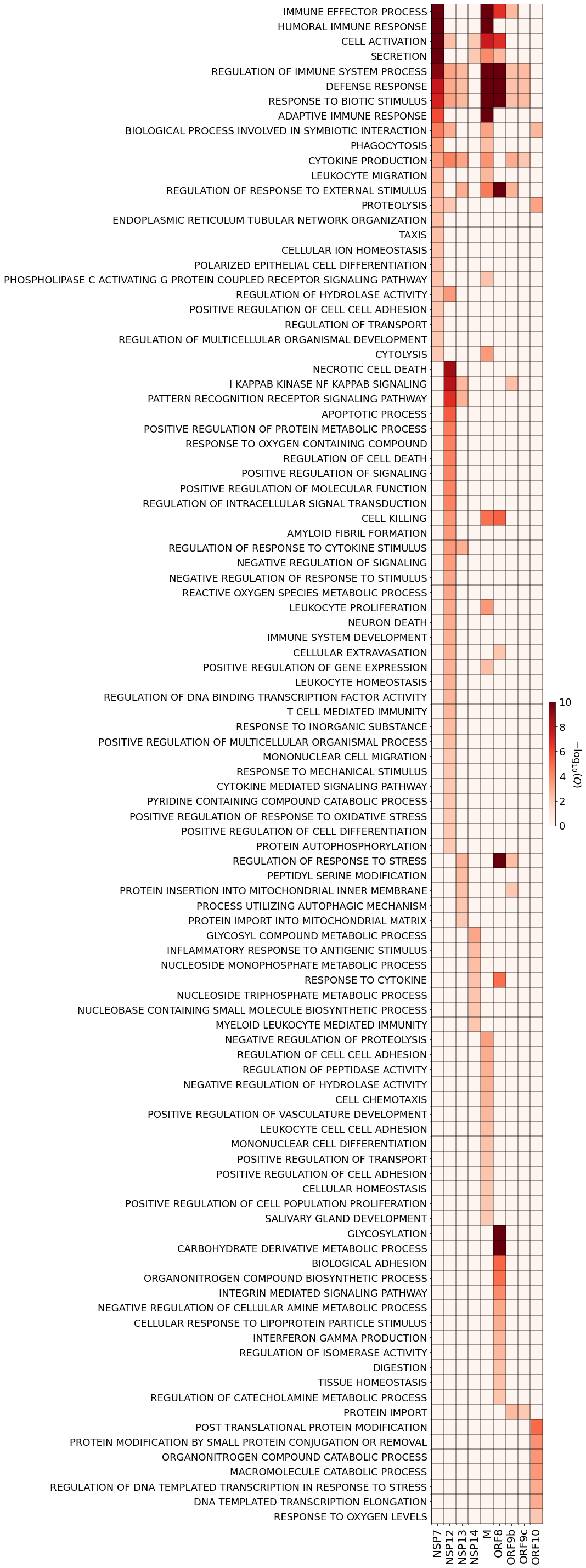} 
    \caption{All 102 significant GOBP terms for the sum of PPI and communicated set of proteins under Protocol 1 for each SARS-CoV-2 protein.}
    \label{fig: all maxEnt go}
\end{figure*}

\begin{figure*}[!ht]
    \includegraphics[height=\textheight]{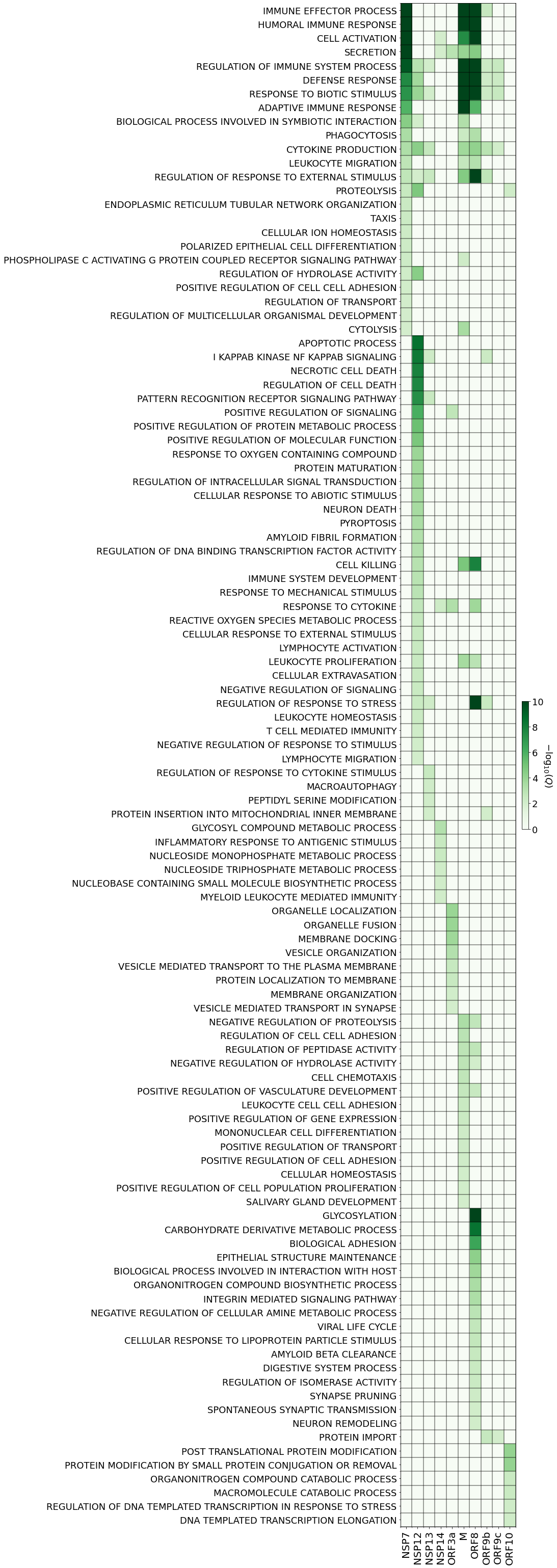} 
    \caption{All 110 significant GOBP terms for the sum of PPI and communicated set of proteins under Protocol 2 for each SARS-CoV-2 protein.}
    \label{fig: all high go}
\end{figure*}

\begin{figure*}[!ht]
    \includegraphics[height=\textheight]{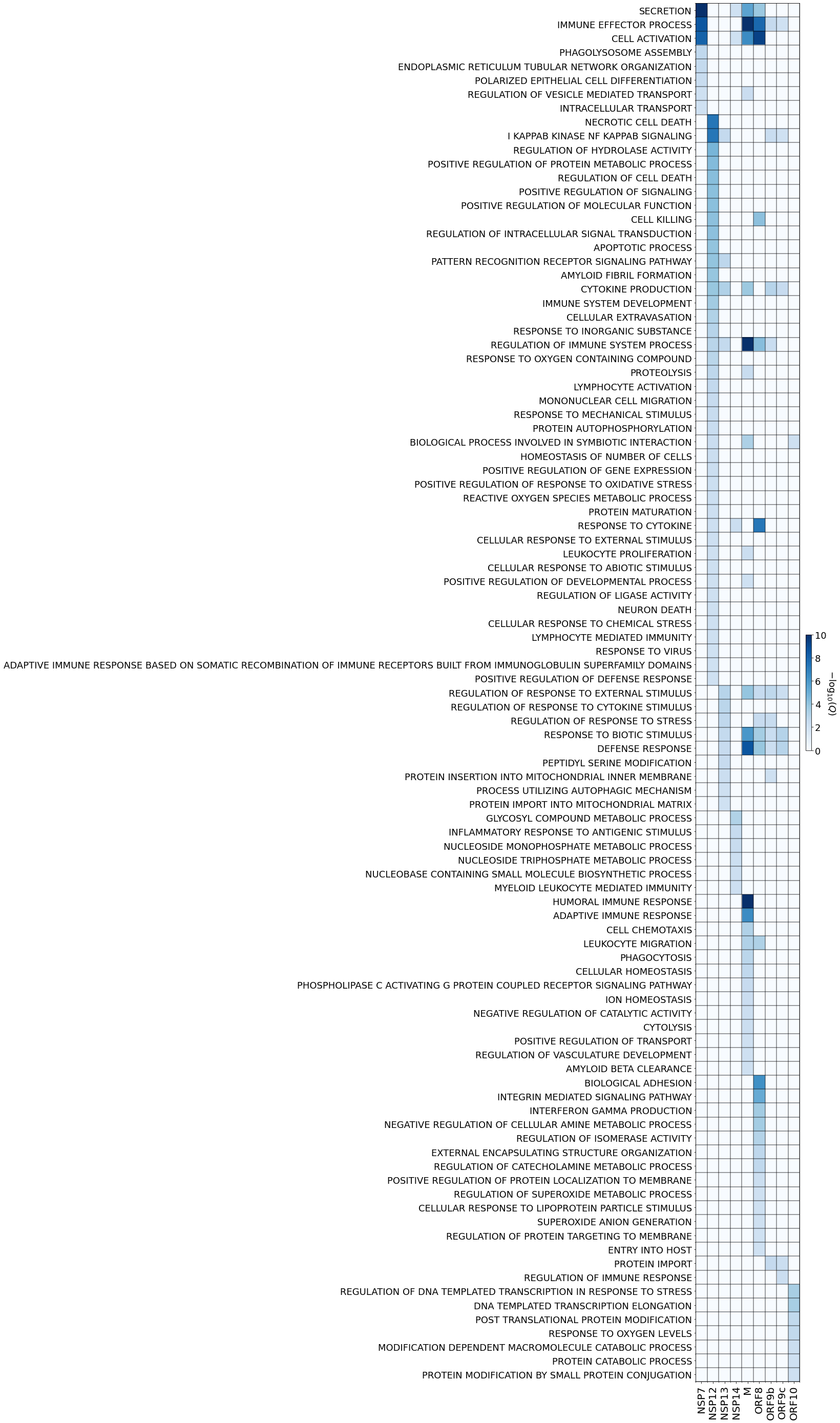} 
    \caption{All 99 significant GOBP terms for the sum of PPI and communicated set of proteins under Protocol 3 for each SARS-CoV-2 protein.}
    \label{fig: all low go}
\end{figure*}

For the figures in the main text (Figures 2 and 3), we identified the top 3 GOBP terms for Protocol 2 for each SARS-CoV-2 proteins, because Protocol 2 produces the maximum network relative entropy due to SARS-CoV-2 PPIs. Figures 2 and 3 in the main text show the adjusted $p$-values for these selected set of GOBP terms only. The full list of all significant GOBP terms for each protocol are in Figures \ref{fig: all maxEnt go}-\ref{fig: all low go}. 

\clearpage

\section{\label{sec: en-ORA}Significant biological processes for a finite number of responses}

To estimate \textit{when} a biological process may become a candidate for dysregulation due to SARS-CoV-2 PPIs, we considered the effect of finite responses and large deviations theory. We present this method for Protocol 1 results, where the reference state for each protein is $\{0.5,0.5\}$, but this analysis can be performed for other reference states too \cite{varadhan1984large,cover1999elements}. As mentioned in \ref{sec: communicated proteins}, we consider the proteins $v_i$ that have a relative entropy $H_{\mathrm{ref}}(v_i) \geq 0.01$ bits to be part of the communicated set. The sum of the PPI and the communicated set is the total interaction set. A typical outcome of comparing the total interaction set against a particular BP gene set is shown in \ref{table: gene BP comparison}.

\begin{table}[h!]
\centering
\begin{tabular}{ c|c|c|c } 
\hline
 & Genes in BP & Genes not in BP & Row total \\ 
\hline
Total interaction set genes & $a=a_{\mathrm{PPI}} + a_{\mathrm{communicated}}$ & $b$ & $a+b$ \\ 
Reference state genes & $c$ & $d$ & $c+d$ \\
\hline
 & $\underline{a+c}$ & $\underline{b+d}$ & $a+b+c+d=\underline{M}$ \\
\end{tabular}
\caption{Typical outcome of comparing the total interaction set proteins with a gene ontology biological process (BP) gene set. The underlined variables are constant, which includes $M$ = the total number of genes across all biological processes; $a+c$ = genes in the chosen biological process; and $b+d$ = genes not in the chosen biological process. The cardinality of the intersection set of a BP gene set and the total intersection set is $a$.}
\label{table: gene BP comparison}
\end{table}
The $p$-value as a function of the variables in Table. \ref{table: gene BP comparison}, which is the probability of obtaining an intersection greater than or equal to $a$ by chance, is given by the hypergeometric survival function as

\begin{equation}
\label{eq: p_value}
p\text{-value} = S_{\text{hg}}(a-1, M, \underline{a+c}, a+b)
\end{equation}
where $S_{\text{hg}}(\cdot)$ is the hypergeometric survival function.

If a protein is in the reference state $\{0.5,0.5\}$ and we sample or draw responses of the protein state for $N$ times, then the empirical distribution $\{P(v_i^{(N)}=1),P(v_i^{(N)}=0)\}$ can be deviated from the reference state just because of stochasticity. So, the relative entropy of the empirical state, $H_{\mathrm{ref}}(v_i^{(N)})$, may be greater than zero, without receiving any communication from SARS-CoV-2 PPIs. The probability of observing such a deviation can be determined using Sanov's theorem, which involves solving a constrained relative entropy minimization problem \cite{cover1999elements}. The solution for a reference state of $\{0.5,0.5\}$ is particularly simple: the probability of obtaining an empirical state after $N$ responses that has a relative entropy greater than or equal to $H_{\mathrm{ref}}(v_i)$ is $2^{-H_{\mathrm{ref}}(v_i)N}$. For example, NSP14 communicates to 5 proteins under Protocol 1 (Table. \ref{table: protocol 1 proteins}). In the absence of SARS-CoV-2 PPIs, the probability of the relative entropy of the empirical state to be $H_{\mathrm{ref}}(v_i)$ or more is in Table \ref{table: finite responses effect}.

\begin{table}[h!]
\sffamily
\small
\centering
\begin{tabular}{ l|l|c|c|c|c|c } 
\hline
& & \sffamily{IL36RN} & IL1F10 & IL1RL2 & IL1RAPL1 & IMPDH1 \\ 
\hline
& $H_{\mathrm{ref}}(v_i)$ & 0.024 & 0.259 & 0.041 & 0.259 & 0.212  \\
\hline
& $N=10$ & 0.85 & 0.17 & 0.75 & 0.17 & 0.23 \\
Pr$[H_{\mathrm{ref}}(v_i^{(N)})\geq H_{\mathrm{ref}}(v_i)]$ & $N=100$ & 0.19 & 0 & 0.06 & 0 & 0 \\
& $N=1000$ & 0 & 0 & 0 & 0 & 0 \\
\hline
\end{tabular}
\caption{Probability of the empirical distribution to have a relative entropy $\geq H_{\mathrm{ref}}(v_i)$ when the protein is actually in the maximum entropy state. The top row of the table lists the 5 proteins that receive communication from NSP14. Pr$[H_{\mathrm{ref}}(v_i^{(N)})\geq H_{\mathrm{ref}}(v_i)]$ values less than 1E-5 are reported as zeros. These probability values quantify the uncertainty of whether a protein has received communication from SARS-CoV-2 PPIs after $N$ responses or we are observing a stochastic deviation.}
\label{table: finite responses effect}
\end{table}

\begin{figure*}[!ht]
    \includegraphics[width=0.44\textwidth]{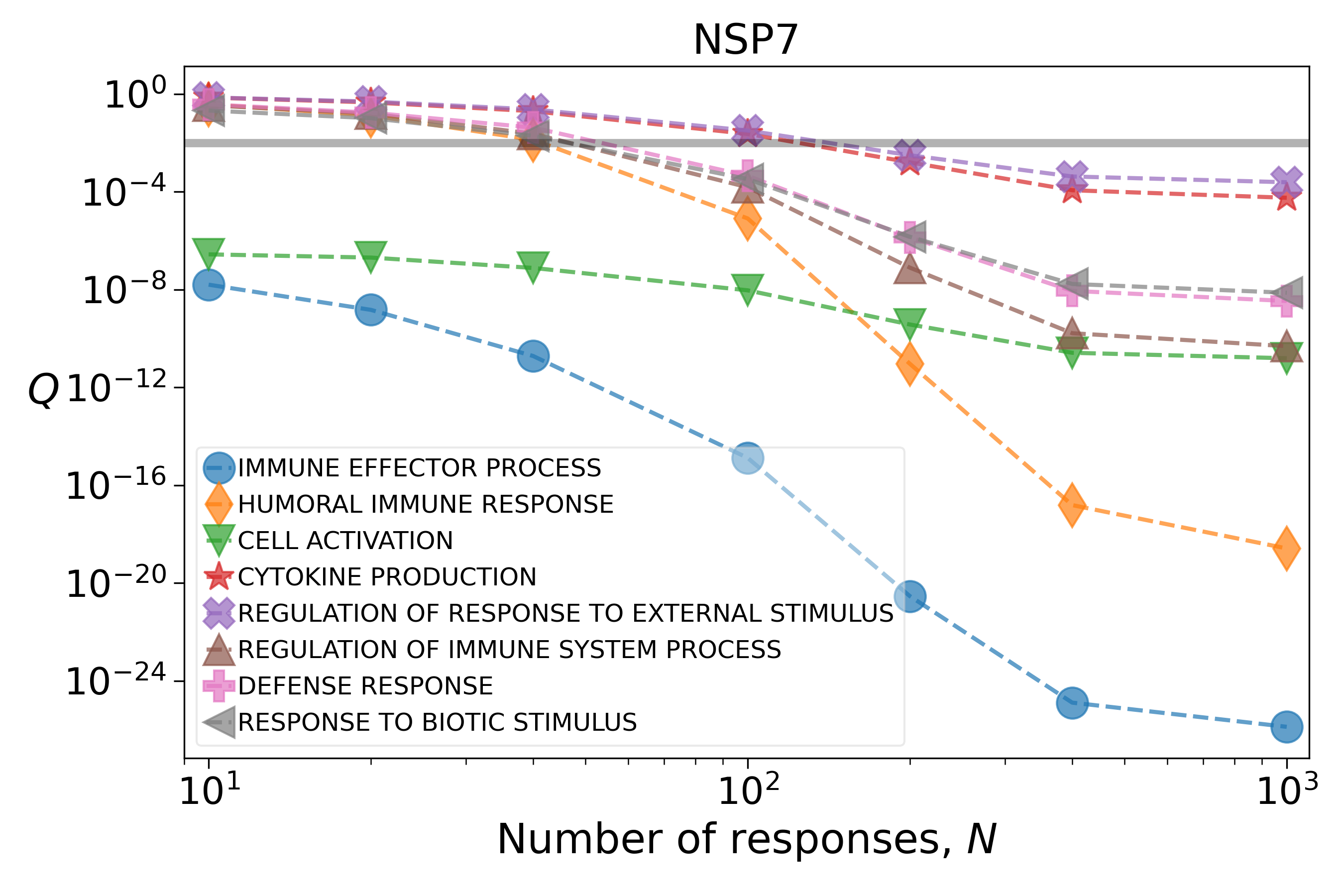} 
    \includegraphics[width=0.44\textwidth]{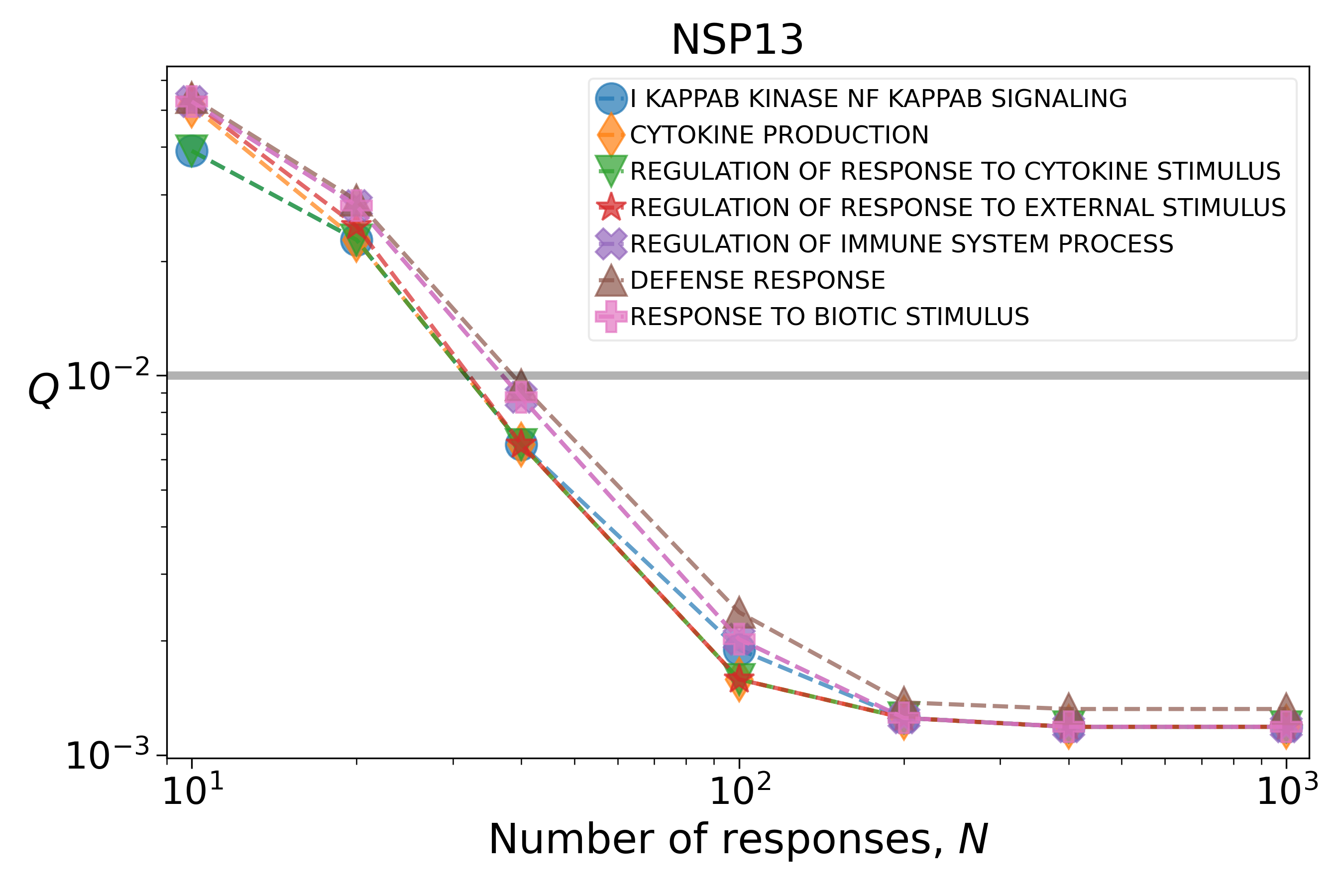} 
    \\
    \includegraphics[width=0.44\textwidth]{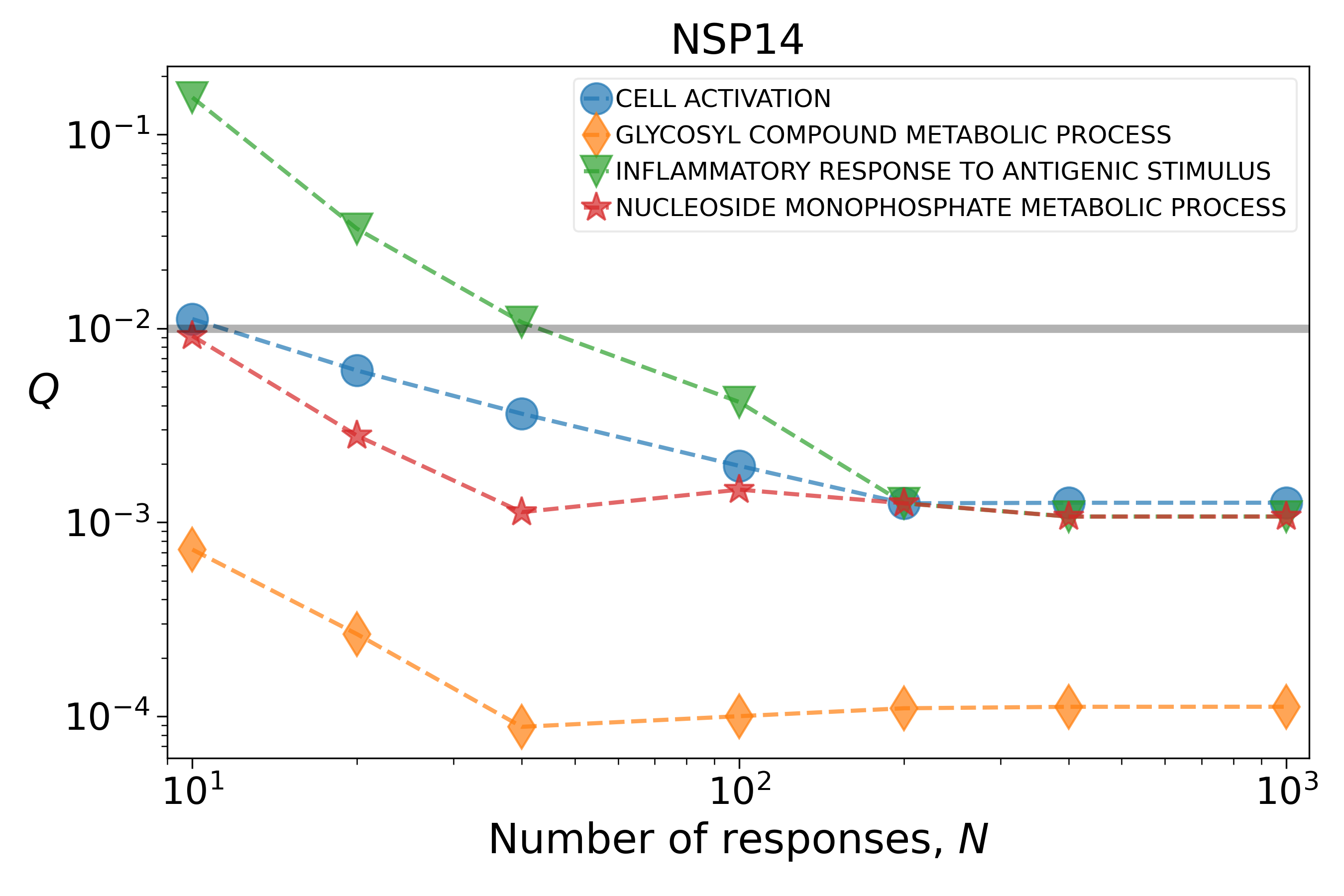} 
    \includegraphics[width=0.44\textwidth]{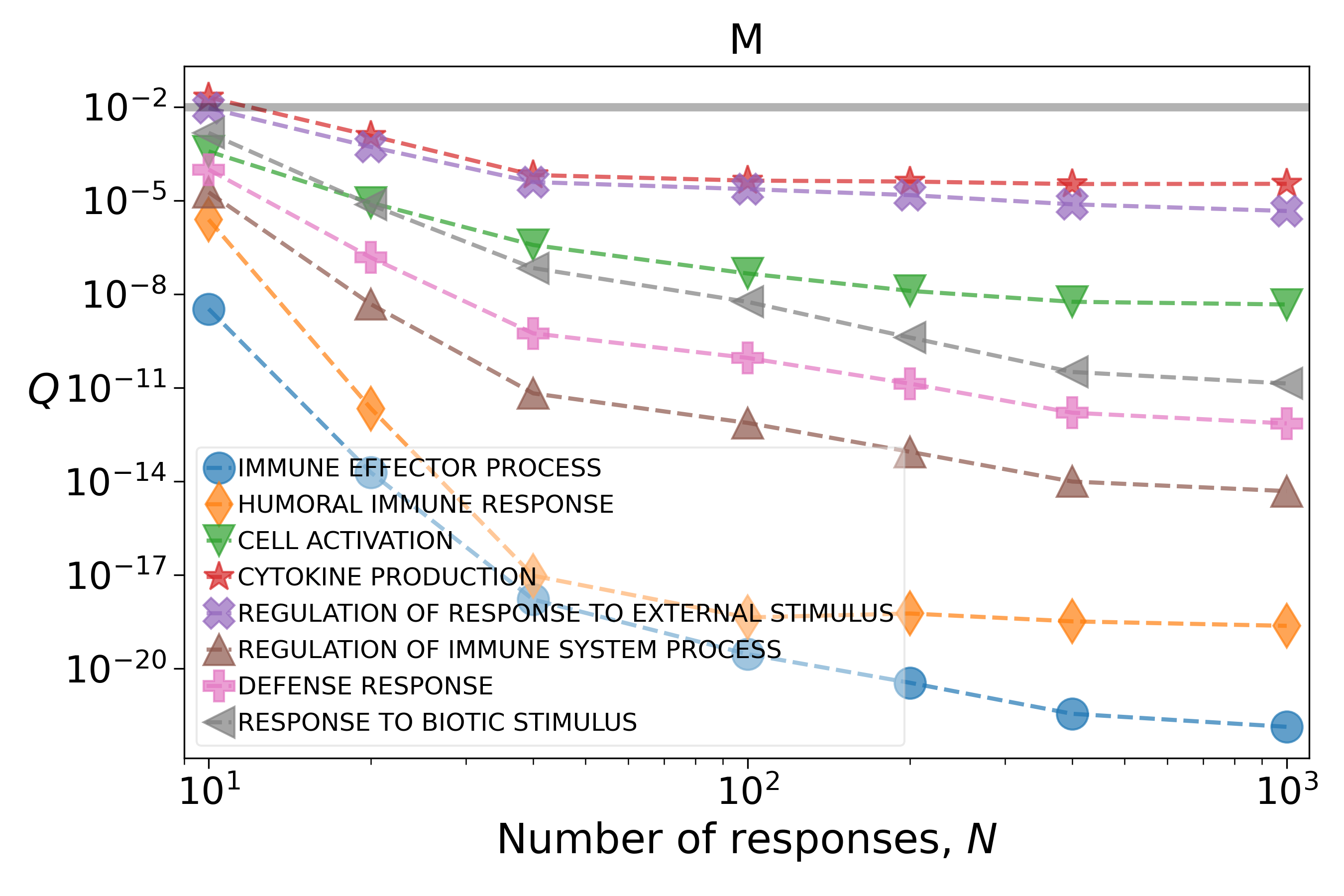}
    \\
    \includegraphics[width=0.44\textwidth]{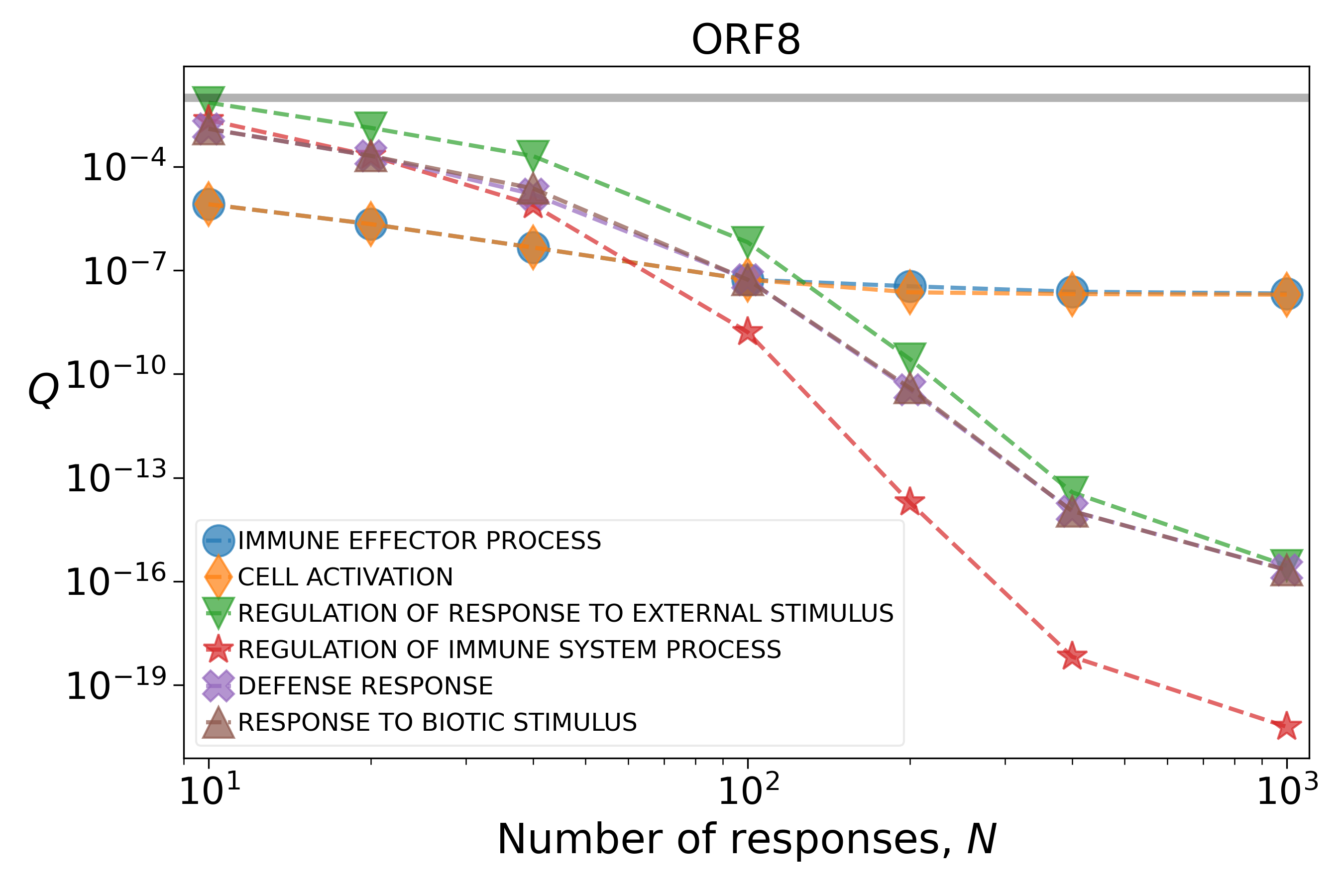} 
    \includegraphics[width=0.44\textwidth]{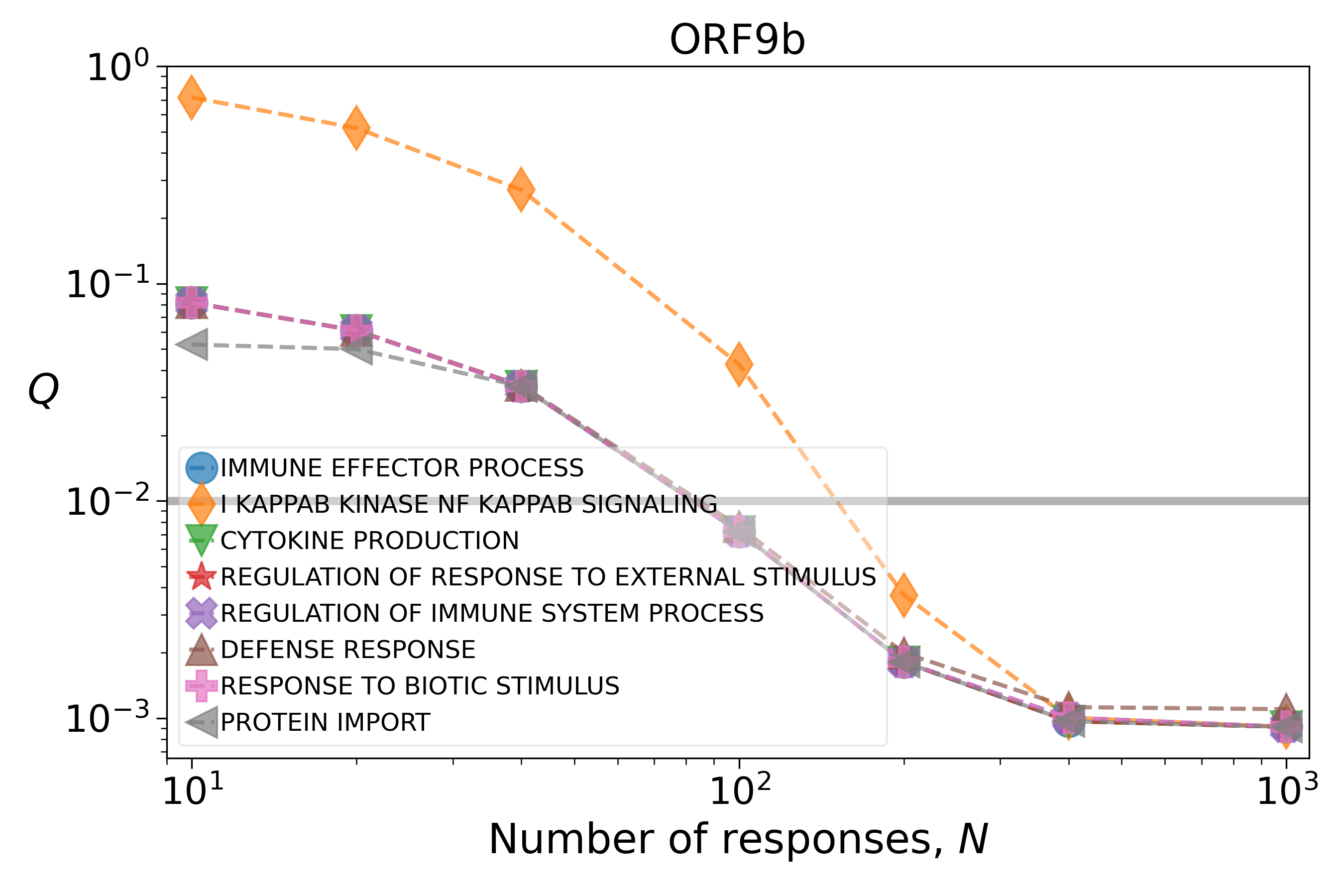}
    \\
    \includegraphics[width=0.44\textwidth]{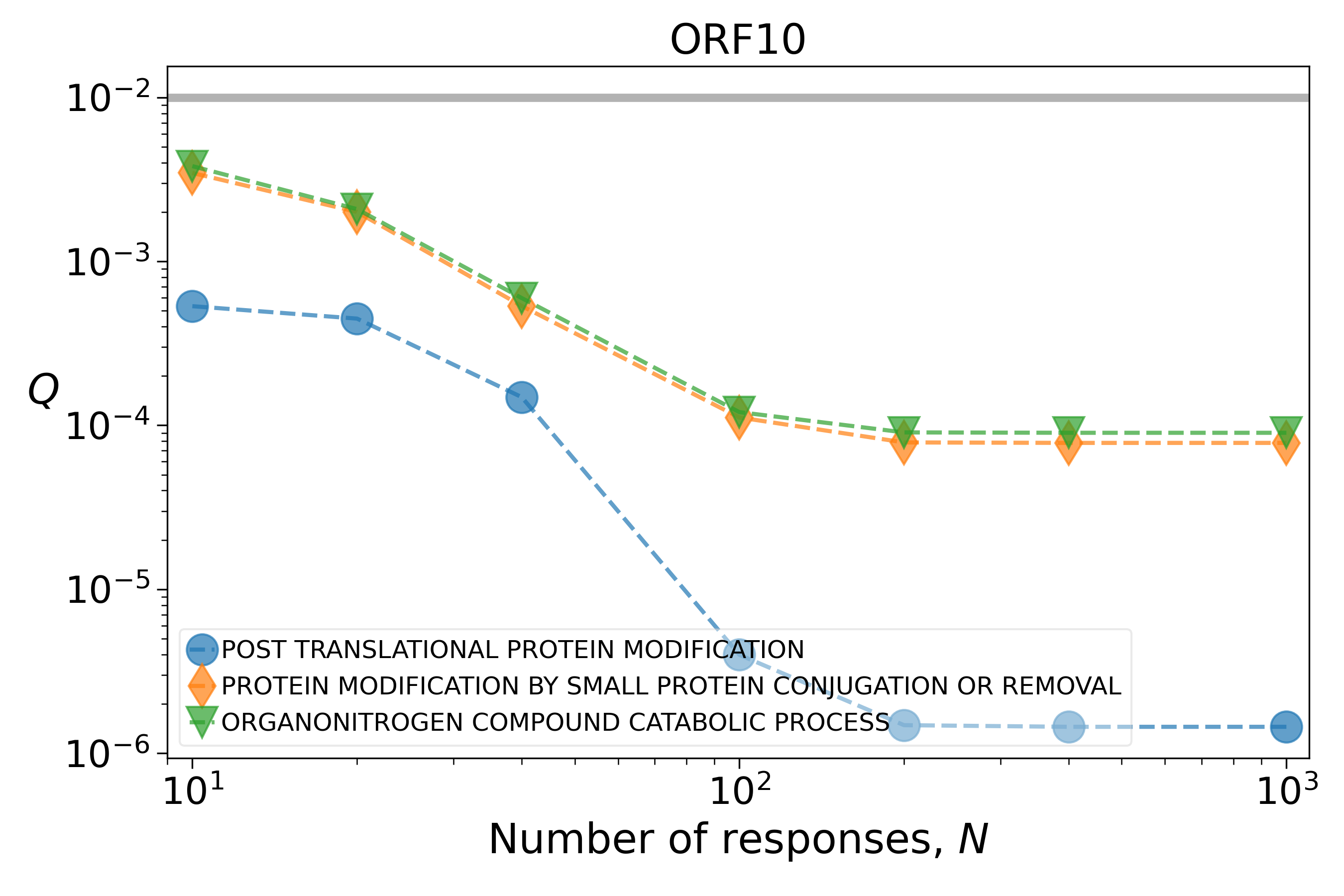} 
    \caption{$Q$-$N$ curves for the significant BPs associated with each SARS-CoV-2 protein. The gray horizontal line in each subfigure show the false discovery rate(1\%) used to identify the significant BPs in Fig. 2 of the main text. (a) Significant BPs for SARS-CoV-2 NSP7 protein. (b) Significant BPs for SARS-CoV-2 NSP13 protein. (c) Significant BPs for SARS-CoV-2 NSP14 protein. (d) Significant BPs for SARS-CoV-2 M (membrane) protein. (e) Significant BPs for SARS-CoV-2 ORF8 protein. (f) Significant BPs for SARS-CoV-2 ORF9b protein. (g) Significant BPs for SARS-CoV-2 ORF10 protein.}
    \label{fig: significant BPs N}
\end{figure*}

To determine the significant biological processes as a function of $N$, we performed a Monte Carlo (MC) simulation to determine the $N$-dependent $Q$ values for the GOBP terms identified in \ref{sec: ORA} under Protocol 1. For each MC sample, we eliminated each protein $v_i$ from the communicated set with the probability $2^{-H_{\mathrm{ref}}(v_i)N}$, which means the total interaction set was itself a random variable corresponding to each MC sample. Subsequently, the outcome of comparing a sample total interaction set with a BP gene set (Table \ref{table: gene BP comparison}) results in random variables $\tilde{a},\tilde{b},\tilde{c}$ and $\tilde{d}$ (still satisfying the constraints mentioned in Table \ref{table: gene BP comparison}). So, from $n_S$ MC samples we computed the average $p$-value as
\begin{equation}
\label{eq: MC p}
\bar{p}\text{-value} = \frac{1}{n_S} \sum_{k=1}^{n_S}S_{\text{hg}}(\tilde{a}^{(k)}-1, M, \underline{a+c}, \tilde{a}^{(k)}+\tilde{b}^{(k)})
\end{equation}
The average $p$-value from Eq. \eqref{eq: MC p} was then used to identify the significant biological processes after Benjamini-Hochberg correction with a false-discovery rate of 1\%. The $Q$ values in Fig. 3 of the main text and in Fig. \ref{fig: significant BPs N} were computed using this procedure. Fig. 3(b) shows the $Q\text{-}N$ curves for the proteins NSP12 and ORF9c, respectively. The plots for the other SARS-CoV-2 proteins are in Fig. \ref{fig: significant BPs N}. We used $n_S=10^4$ for this calculation and verified for convergence using $n_s=4\times 10^4$. 

\clearpage

\bibliography{references}